\documentclass[TOTEM]{cernphprep}
\usepackage{graphicx}
\DeclareGraphicsExtensions{.pdf,.png,.gif,.jpg,.mht}
\graphicspath{{./Fig-timing/}}
\usepackage[noadjust]{cite}
\usepackage{multirow}
\usepackage{amssymb}
\usepackage{verbatim}
\usepackage{hyperref}

\usepackage{lineno}
\usepackage{geometry}
\usepackage{topcapt} 
\def\etal{et al.}

\def\Instline#1#2{%
	\expandafter\write1{\string\newlabel{#1}{{#1}{}}}%
	\hbox to\hsize{\strut\hss$^{#1}$#2\hss}
}

\begin{document}

\begin{titlepage}

\PHnumber{2016-317}
\PHdate{ 17 December 2016}
\EXPnumber{TOTEM 2016--05}
\EXPdate{17 December 2016}
\DEFCOL{CDS-Library}

\title{Diamond Detectors for the TOTEM Timing Upgrade}
\ShortTitle{TOTEM timing detectors}

\Collaboration{The TOTEM Collaboration}
\ShortAuthor{The TOTEM Collaboration (G.~Antchev \emph{\etal})}

\begin{Authlist}
G.~Antchev\Aref{a},
P.~Aspell\Iref{9},
I.~Atanassov\Aref{a},
V.~Avati\Iref{8},
J.~Baechler\Iref{9},
V.~Berardi\IIref{5b}{5a},
M.~Berretti\IIref{3a}{9},
E.~Bossini\Iref{7b},
U.~Bottigli\Iref{7b},
M.~Bozzo\Iref{6a},
P.~Broul\'{i}m\Iref{1a},
A.~Buzzo\Iref{6a},
F.~S.~Cafagna\Iref{5a},
M.~G.~Catanesi\Iref{5a},
M.~Csan\'{a}d\IAref{4a}{b},
T.~Cs\"{o}rg\H{o}\IIref{4a}{4b},
M.~Deile\Iref{9},
F.~De~Leonardis\IIref{5c}{5a},
A.~D'Orazio\IIref{5c}{5a},
M.~Doubek\Iref{1c},
K.~Eggert\Iref{10},
V.~Eremin\Aref{d},
F.~Ferro\Iref{6a},
A. Fiergolski\Iref{9},
F.~Garcia\Iref{3a},
V.~Georgiev\Iref{1a},
S.~Giani\Iref{9},
L.~Grzanka\IAref{8}{c},
C.~Guaragnella\IIref{5c}{5a},
J.~Hammerbauer\Iref{1a},
J.~Heino\Iref{3a},
A.~Karev\Aref{e}
J.~Ka\v{s}par\IIref{1b}{9},
J.~Kopal\Iref{1b},
V.~Kundr\'{a}t\Iref{1b},
S.~Lami\Iref{7a},
G.~Latino\Iref{7b},
R.~Lauhakangas\Iref{3a},
R.~Linhart\Iref{1a},
M.~V.~Lokaj\'{\i}\v{c}ek\Iref{1b},
L.~Losurdo\Iref{7b},
M.~Lo~Vetere\IIref{6b}{6a},
F.~Lucas~Rodr\'{i}guez\Iref{9},
D.~Lucs\'anyi\Iref{9},
M.~Macr\'{\i}\Iref{6a},
A.~Mercadante\Iref{5a},
N.~Minafra\IIref{9}{5b},
S.~Minutoli\Iref{6a},
T.~Naaranoja\IIref{3a}{3b}
F.~Nemes\IAref{4a}{b},
H.~Niewiadomski\Iref{10},
T.~Nov\'ak\IIref{4a}{4b},
E.~Oliveri\Iref{9},
F.~Oljemark\IIref{3a}{3b},
M.~Oriunno\Aref{f},
K.~\"{O}sterberg\IIref{3a}{3b},
P.~Palazzi\Iref{9},
L.~Palo\v{c}ko\Iref{1a},
V.~Passaro\IIref{5c}{5a},
Z.~Peroutka\Iref{1a},
V.~Petruzzelli\IIref{5c}{5a},
T.~Politi\IIref{5c}{5a},
J.~Proch\'{a}zka\Iref{1b},
F.~Prudenzano\IIref{5c}{5a},
M.~Quinto\Iref{9},
E.~Radermacher\Iref{9},
E.~Radicioni\Iref{5a},
F.~Ravotti\Iref{9},
E.~Robutti\Iref{6a},
C.~Royon\Aref{g},
G.~Ruggiero\Iref{9},
H.~Saarikko\IIref{3a}{3b},
A.~Scribano\Iref{7a},
J.~Smajek\Iref{9},
W.~Snoeys\Iref{9},
J.~Sziklai\Iref{4a},
C.~Taylor\Iref{10},
N.~Turini\Iref{7b},
V.~Vacek\Iref{1c},
J.~Welti\IIref{3a}{3b},
P.~Wyszkowski\Iref{8},
K.~Zielinski\Iref{8}
\end{Authlist}

\Instline{1a}{University of West Bohemia, Pilsen, Czech Republic.}
\Instline{1b}{Institute of Physics of the Academy of Sciences of the Czech Republic, Praha, Czech Republic.}
\Instline{1c}{Czech Technical University, Praha, Czech Republic.}
\Instline{2}{National Institute of Chemical Physics and Biophysics NICPB, Tallinn, Estonia.}
\Instline{3a}{Helsinki Institute of Physics, Helsinki, Finland.}
\Instline{3b}{Department of Physics, University of Helsinki, Helsinki, Finland.}
\Instline{4a}{Wigner Research Centre for Physics, RMKI, Budapest, Hungary.}
\Instline{4b}{EKE KRC, Gy\"ongy\"os, Hungary.}
\Instline{5a}{INFN Sezione di Bari, Bari, Italy.}
\Instline{5b}{Dipartimento Interateneo di Fisica di Bari, Bari, Italy.}
\Instline{5c}{Dipartimento di Ingegneria Elettrica e dell'Informazione - Politecnico di Bari, Bari, Italy.}
\Instline{6a}{INFN Sezione di Genova, Genova, Italy.}
\Instline{6b}{Universit\`{a} degli Studi di Genova, Italy.}
\Instline{7a}{INFN Sezione di Pisa, Pisa, Italy.}
\Instline{7b}{Universit\`{a} degli Studi di Siena and Gruppo Collegato INFN di Siena, Siena, Italy.}
\Instline{8}{AGH University of Science and Technology, Krakow, Poland.}
\Instline{9}{CERN, Geneva, Switzerland.}
\Instline{10}{Case Western Reserve University, Dept. of Physics, Cleveland, OH, USA.}

\Anotfoot{a}{INRNE-BAS, Institute for Nuclear Research and Nuclear Energy, Bulgarian Academy of Sciences, Sofia, Bulgaria.}
\Anotfoot{b}{Department of Atomic Physics, ELTE University - Budapest, Hungary}
\Anotfoot{c}{Institute of Nuclear Physics, Polish Academy of Science, Krakow, Poland.}
\Anotfoot{d}{Ioffe Physical - Technical Institute of Russian Academy of Sciences, St. Petersburg, Russian Federation}
\Anotfoot{e}{JINR, Dubna, Russia.}
\Anotfoot{f}{SLAC National Accelerator Laboratory, Stanford CA, USA.}
\Anotfoot{g}{University of Kansas, Lawrence, KS, USA.}

\begin{abstract}
This paper describes the design and the performance of the timing detector developed by the TOTEM Collaboration for  the Roman Pots (RPs) to measure the Time-Of-Flight (TOF) of the protons produced in central diffractive interactions at the LHC. 
The measurement of the TOF of the protons allows the determination of the  longitudinal position of the proton interaction vertex and its association  with one of the vertices reconstructed by the CMS detectors. 
The TOF detector is based on single crystal Chemical Vapor Deposition (scCVD) diamond plates and is designed to measure the protons TOF with about 50 ps time precision. 
This upgrade to the TOTEM apparatus will be used in the LHC run 2 and will tag the central diffractive events  up to an interaction pileup of about 1.   
A dedicated fast and low noise electronics for the signal amplification has been developed.
The digitization of the diamond signal is performed by sampling the waveform. 
After introducing the  physics studies that will most profit  from the addition of these new detectors, we discuss in detail the optimization and the performance of the first TOF detector installed in the LHC in November 2015.

\end{abstract}


\end{titlepage}


\tableofcontents

\newpage

\section{Introduction}

The TOTEM experiment at CERN's Large Hadron Collider (LHC) was optimized for the measurement of the elastic pp scattering over a four-momentum transfer squared $|t|$ ranging ultimately from $\le 10^{-3}\,\rm GeV^{2}$ to $\sim 10\,\rm GeV^{2}$ and has measured the total pp cross-section in dedicated special-optics runs~\cite{Antchev:2011vs,Antchev:2013iaa,totem4,PhysRevLett.111.012001} and studied the inelastic rate~\cite{Antchev:2013iaa,totem5}.

Diffractive scattering represents a unique tool for investigating the dynamics of strong interactions and proton structure. These events are dominated by soft processes which cannot be calculated with perturbative QCD. 
Various model calculations predict diffractive cross-sections that are markedly different at the LHC energies~\cite{Ryskin:2007qx,Gotsman:2008tr,Ostapchenko:2011nk}.
 A special optics configuration of the accelerator (large $\beta ^*$) permits to see all masses in central diffractive topologies. 
 
A recently approved upgrade of the TOTEM detectors~\cite{CERN-LHCC-2014-024} will add the possibility to measure the arrival time of the leading protons in the Roman Pots (RP).
Among many physics channels the upgraded system will provide an unprecedented sensitivity for low mass resonances (with particular emphasis on glueball candidates), exclusive Central Diffractive (CD) dijets, charmonium states and events with missing mass signatures~\cite{Osterberg:2014mta}. 
Even at very low pileup, for inclusive CD events and in particular for events with missing momentum the association of the protons to the CMS vertex by using only the tracking variables is problematic. 
Here, as well as in all CD events measured at moderate pileup, the proton Time-Of-Flight (TOF) measurement becomes crucial. 
As an example, at $\mu=$50\%\,\footnote{$\mu$ is defined as the average number of inelastic proton-proton interactions per bunch crossing.}, a  five-fold enhancement  of the inclusive CD  purity can be obtained from the installation of the TOF detector in the RP, assuming a  time precision of 50 ps~\cite{CERN-LHCC-2014-020}. 

\section{The TOTEM Apparatus}
\subsection{Overview}
The TOTEM experiment~\cite	{TOTEM-TDR} is composed of three sets of subdetectors placed symmetrically on both sides of the interaction point: the Roman Pot detectors identify leading protons whereas the T1 and T2 telescopes detect charged particles in the forward region from proton diffractive dissociation. 
T2 consists of Gas Electron Multipliers that detect charged particles with $p_{T}>\,$40~MeV/c at pseudo-rapidities of 5.3$< |\eta|<$6.5. 
The T1 telescope consists of Cathode Strip Chambers that measure charged particles with $p_{T}>\,$100~MeV/c at 3.1$<$ $|\eta|$ $<$4.7. 
T1 and T2 can be used to tag rapidity gaps in CD events (p + p $\rightarrow$ p + X + p).
At present, to detect leading protons scattered at very small angles, silicon sensors are placed in movable beam-pipe insertions -- so-called ``Roman Pots" (RP), special sections of the LHC vacuum chamber that can be positioned very close  to the circulating beam~\cite{Anelli:2008zza}. 
In order to maximize the experiment's acceptance for elastically scattered protons, the RP are able to approach the beam centre to a transverse distance as small as 1\,mm or, equivalently, to a few $\sigma_{beam}$ (the transverse beam size).

Each RP station is composed of two units separated by a distance of about 5\,m. A unit consists of three RPs, two approaching the outgoing beam vertically and one horizontally.  
The detectors in the horizontal pot complete the acceptance for diffractively scattered protons.
All RPs are rigidly fixed within the unit, together with a Beam Position Monitor (BPM).

\subsection{The Existing Roman Pot Tracking Detectors}

Each RP is equipped with a stack of 10 silicon strip detectors designed with the specific objective of reducing to only a few tens of micrometers~\cite{Ruggiero:2009zz} the inefficient area at the detector edge closest to the beam.
The 512 strips with 66\,$\mu$m pitch of each detector are oriented at an angle of $+ 45^\circ$ (five planes) and $- 45^\circ$ (five planes) with respect to the detector edge facing the beam.
During the measurement the detectors in the horizontal RPs overlap with the ones in the vertical RPs, enabling a precise (10\,$\mu $m) relative alignment of the detectors in the three RPs of a unit by correlating their positions via common particle tracks.
These detectors are described in detail in ~\cite{Anelli:2008zza}.
To provide multi track separation capability TOTEM has added in 2014 in each arm one RP unit tilted in the vertical plane by $8^{\circ}$.
 
\subsection{TOF Measurement in TOTEM}

In order to improve the experiment's capability to explore and measure new physics in CD processes TOTEM has proposed an upgrade program ~\cite{CERN-LHCC-2014-024} which foresees adding timing capability to the proton detectors installed in the very forward region.
Common CMS-TOTEM data taking is foreseen during the LHC Run 2, with a special LHC-optics configuration for which the proton acceptance is optimal (all $\xi = \Delta$p/p for $|t|\,>\,$0.04 GeV$^2$).
The addition of proton Time-Of-Flight (TOF) detectors in the TOTEM forward proton arms allows to reconstruct the longitudinal interaction position and thus to associate this to the proton vertex reconstructed by the CMS tracker. 

The CD interactions measured by TOTEM at $\sqrt{s}=13$ TeV are characterized by having two high energy protons (with momentum  greater than 5 TeV/c) scattered at less than 100 $\mu$rad from the beam axis.
The tracking of the protons is done by silicon strip detectors installed in the RPs.
With the addition of protonTOF detectors, it will be possible to measure the difference of the arrival times, which is directly proportional to the longitudinal position of the vertex.
In fact if $t_1$, $t_2$ are the arrival times of the protons coming from a CD interaction in the two RP arms, the longitudinal vertex position given by $z=c (t_1-t_2)/2$.

During the programmed joint CMS-TOTEM data taking when the two experiments share a common trigger, the reconstruction of the proton vertex position allows, even in presence of event pileup, the association with one of the CD vertex reconstructed by CMS and therefore to completely reconstruct the event.

The timing detector will be installed in four TOTEM vertical Roman Pots (RP) located at $\sim$220~m on both sides of the interaction point 5 (IP5) of the LHC.

\section{The Timing Detectors}

\subsection{Requirements and Design Principles}

\begin{figure}[ht]
 \begin{center}
 \includegraphics[width=0.7\linewidth] {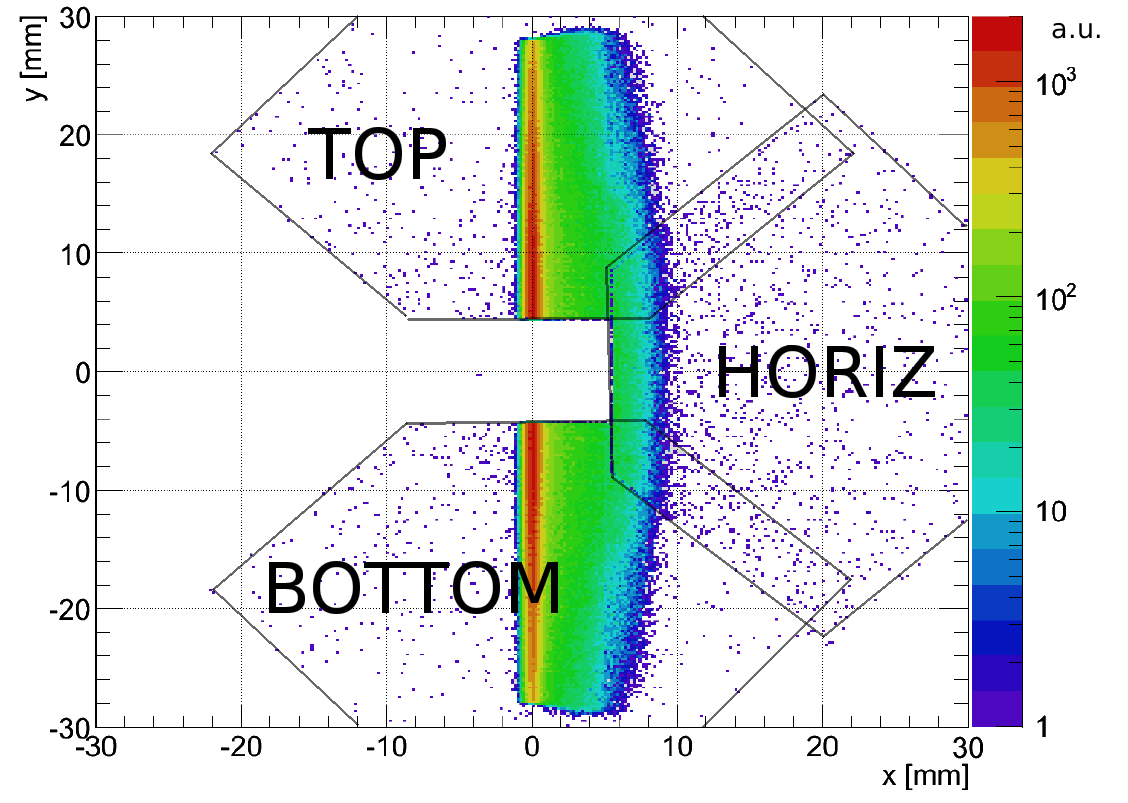}
 \caption{The distribution of the particles traversing the RP plane at 220m with the TOF detector to be installed in top and bottom RP for $\beta ^*$ = 90 m optics. The TOF detectors will be installed in top and bottom RPs.}
  \label{fig:scatter}
 \end{center} 
\end{figure}

The physics program determines the specifications for the timing detector: in order to obtain a precision on the longitudinal interaction vertex ($Z_{vertex} = c\Delta t/2$) of $\sigma_Z$ $\sim$1 cm, an overall time precision $\sigma(\Delta t)\sim$50 ps for the measurement of the proton track timing  is required.
The track distribution at the position of the timing detectors due to diffractive events and to overlapping background is not uniform  and has been measured from RP tracking data, see Figure~\ref{fig:scatter}.
The average number of particles per bunch crossing in one of the RP is about 0.2 (including the beam background) at $\mu = $\,0.5\,\% \cite{Berretti:1747282}. 

The detector has to be segmented into  pixels to minimize the occupancy\footnote{number of particles  traversing the detector cell/pixel in one bunch crossing.} of each cell.

\begin{figure}[ht]
  \begin{center}
 \includegraphics[width=0.9\linewidth]{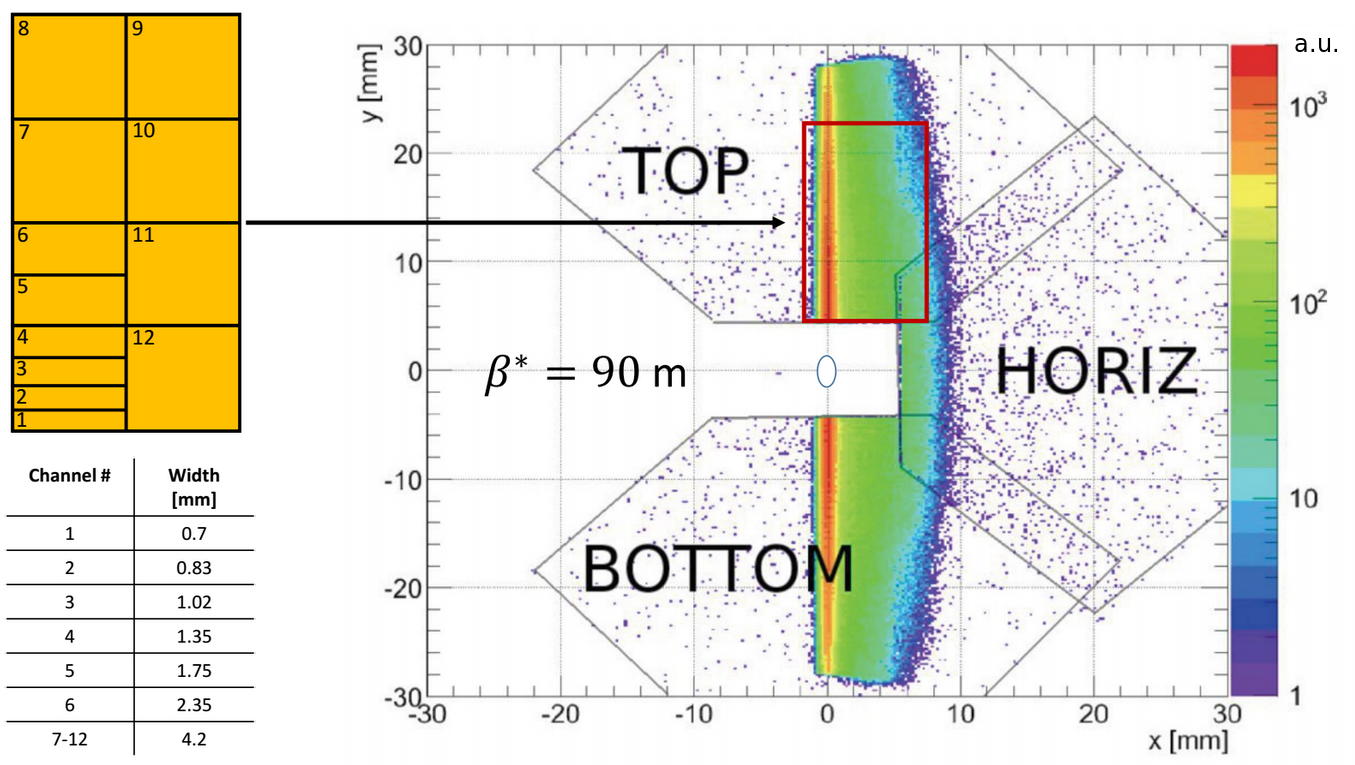}
 \caption{Optimized layout of the detector pixels. On the right its position during physics data taking is outlined on the track distribution of Figure~\ref{fig:scatter} .}
 \label{fig:general-layout}
 \end{center} 
\end{figure}

To guarantee uniform occupancy and a small number of cells  in the detector a pattern with pixels of different sizes has been simulated~\cite{Berretti:1747282} and optimized in order to minimize the inefficiency due to the simultaneous presence of two hits in the same cell. 
As the occupancy in this scenario is anyway very low, this selection hardly affects the efficiency. 
The optimized detector geometry with eight diamond plates obtained for a high-$\beta^*$ optics configuration at $\mu$= 0.5 is shown in Figure~\ref{fig:general-layout}.  
The pixel size in the vertical direction y varies by almost a factor six from 0.7 mm to 4.2~mm. 

At $\mu=$ 50\%, the inefficiency introduced by the pileup is approximately 0.6\%.

\begin{figure}[h!]
  \begin{center}
 \includegraphics[width=0.42\linewidth]{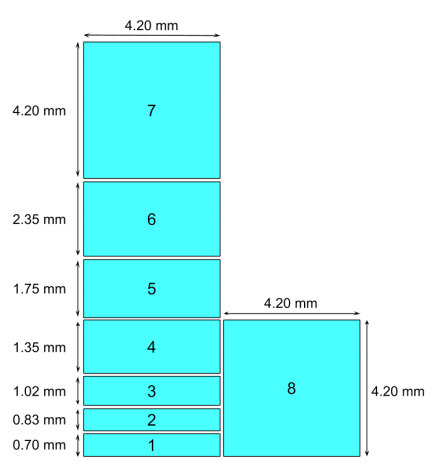}
 \caption{Layout and dimensions of the pixels  of the TOTEM timing detectors implemented and described in this paper.}
 \label{fig:modified-layout}
 \end{center} 
\end{figure}

A reduced geometry has been adopted to build the prototype described here: it contains only four diamond plates and is shown in Figure~\ref{fig:modified-layout}. 
The four diamond single-pixel plates that are not included would cover only an  extra 10\%  of acceptance.

One detector package contains four identical planes, each equipped with four plates of $4.5\times4.5\,{\rm mm}^2$ and  $500\, \mu$m thick scCVD diamond metalized with a pattern of one, two or four pixels.
If one can build a single sensor capable of measuring with a time precision of the order of 100 ps, then with time measurements from four detectors one reaches the required overall time precision  of about 50 ps.
With such a precision for the measurement of the proton track timing the longitudinal interaction vertex can be determined with an uncertainty of $\sim$1 cm.

After some preliminary measurements in a test beam with a diamond sensor assembled with electronics of preliminary  design, we saw that scCVD diamond detectors with a custom-built amplifier would match all the requirements of the timing detector needed for the TOTEM upgrade. 

With an appropriate metalization, one obtains a good contact to the face of the diamond sensor; it is fairly easy (fast and economic) then to implement a pattern with pixels of different sizes on a diamond crystal surface by means of a simple metalization.
Due to the extremely high impedance of the diamond material, the difference in pixel size will barely affect the time response of the signal.
Diamond sensors show also a better radiation hardness~\cite{2007PSSAR.204.3004D} and shorter charge collection times than other semiconductor sensors~\cite{pernegger.2005,Pomorski-thesis}. 


Additional technical requirements were taken into account to design a detector system that could be implemented and properly function in the high radiation environment of the LHC tunnel.
In the following sections we detail the studies performed to ensure that scCVD  diamond detectors will be able to meet the timing precision required by the physics program.

\begin{figure}[h!]
  \begin{center}
\includegraphics[width=0.8\linewidth]{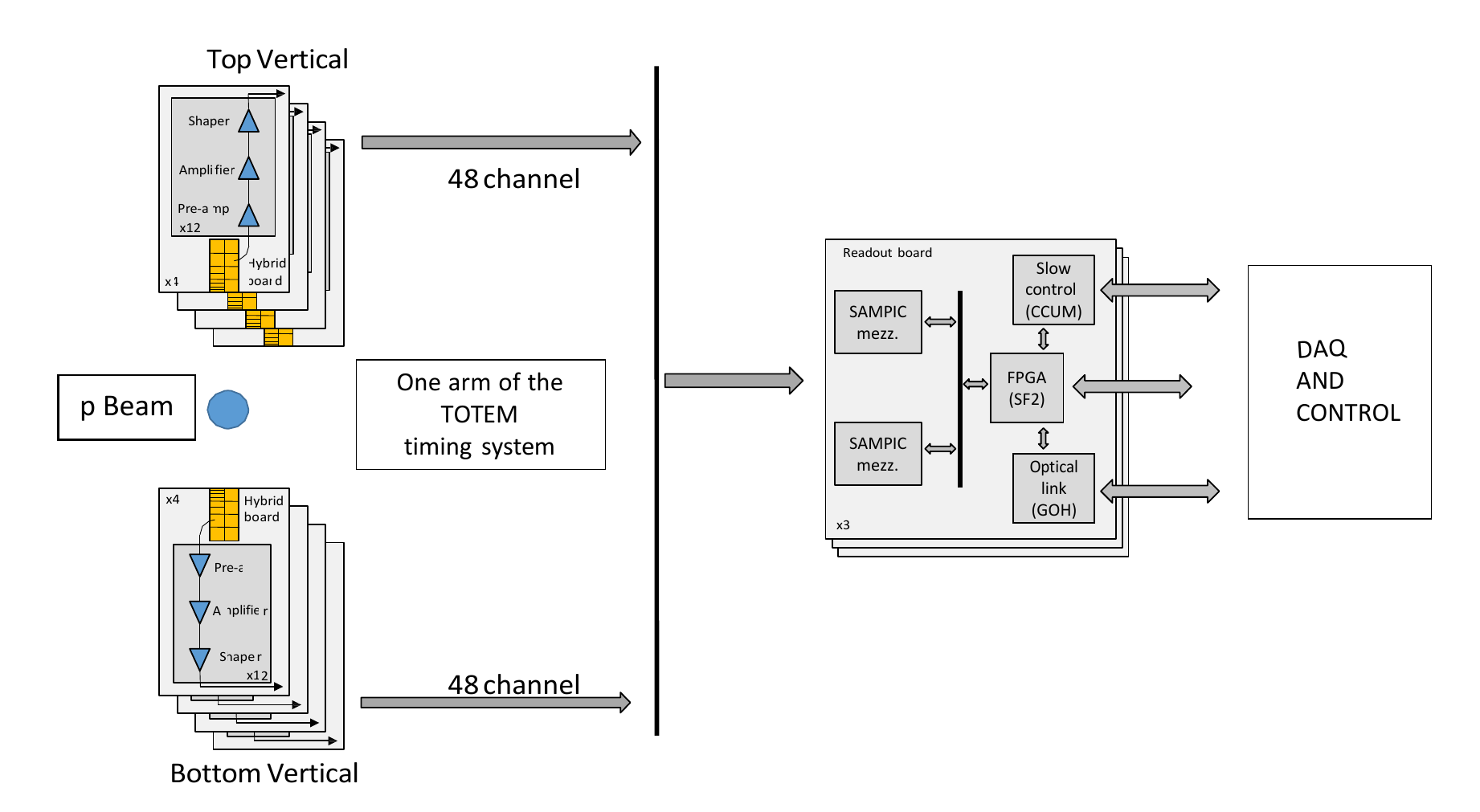}
 \caption{The  TOF system for the TOTEM Vertical RP upgrade.}
  \label{fig:block-diagram}
 \end{center} 
\end{figure}

The TOF system main blocks which will be described in detail in the following sections is shown in Figure~\ref{fig:block-diagram}.

\subsection{The Diamond Sensors}

 The use of detector grade scCVD diamond for use in particle detectors has been proposed in the past and extensively studied~\cite{Pomorski-thesis}.
It is an ideal material for detecting particles due to its high electron and hole mobility and $>$99\% charge collection efficiency.
 
These detectors are now widely used in many specific applications and experiments, albeit still on a small scale. 
The availability on the market of detector grade scCVD diamond detectors of different sizes and thicknesses has greatly improved in recent years.
Several companies\footnote{ \href{http://www.e6.com}{Element Six Ltd.}, King’s Ride Park, Ascot, Berkshire SL5 8BP, UK., \href{http://www.diam2tek.eu}{Diam2Tek}, Irma-Feldweg-Stra{\ss}e 8, 75179 Pforzheim, Germany, among few others.} offer custom sizes and thicknesses  with reasonable delivery time.

\subsection{Diamond Crystal Characterization}\label{sec:diamond}

Electrodes on both sides are needed to collect the charges and read out the electric signal.
Thin metal layers  on both the top and bottom surface of the crystal generate the proper field and allow for connection to the amplifier.
However, making a good ohmic electrical contact with the crystal structure is a delicate process. 
Metalization of the pixels on the surface of the diamond in some instances led to poor collection of charge or to a particle-rate dependent reduction of the charge collection~\cite{Pomorski-thesis}.
The occurrence of `charging' and `radiation' damage of the diamond sensors  has also been studied in high fluence High Energy Physics experiments, see for example~\cite{Guthoff:2013hja}.  

The metalization area for the TOTEM plates covers $4.2\times 4.2\,{\rm mm}^2$ leaving a non-metalized border region of 150$\,\mu$m  in order to avoid enhancement of surface leakage currents and discharges through the diamond edge.
The metalization mask is aligned with better than 100$\mu$m precision on the diamond plate.
When more than one pixel is present on one plate, the non metalized separation between pixels is 100$\mu$m.
The diamond plates (four are needed per hybrid) are glued on the hybrid PCB in contact to each other and one may consider their maximum distance to be less than 50$\mu$m. 

Often the metalization processes are company specific and confidential.
We have tested and compared different metalization techniques for the diamond sensors as suggested and provided by: 
\begin{itemize} \itemsep0pt \parskip0pt \parsep0pt 
\item GSI\footnote{http://www.gsi.de} (Cr-50 nm + Au-150 nm) 
\item Applied Diamond Inc. (USA)\footnote{http://usapplieddiamond.com/} (Cr-50 nm + Au-150 nm) 
\item PRISM\footnote {Micro fabrication laboratory at Princeton University, {http://www.princeton.edu/prism/}} (100 nm of 10\% Ti and 90\% W alloy).
\end{itemize}

After a preliminary optical inspection with the microscope to check for small defects and missing or non-uniform metalization, the TOTEM  high purity $<$100$>$ orientation scCVD diamond plates have been checked  for leakage current and stable signal current.

\begin{figure}[ht]
  \begin{center}
\includegraphics[width=0.8\linewidth]{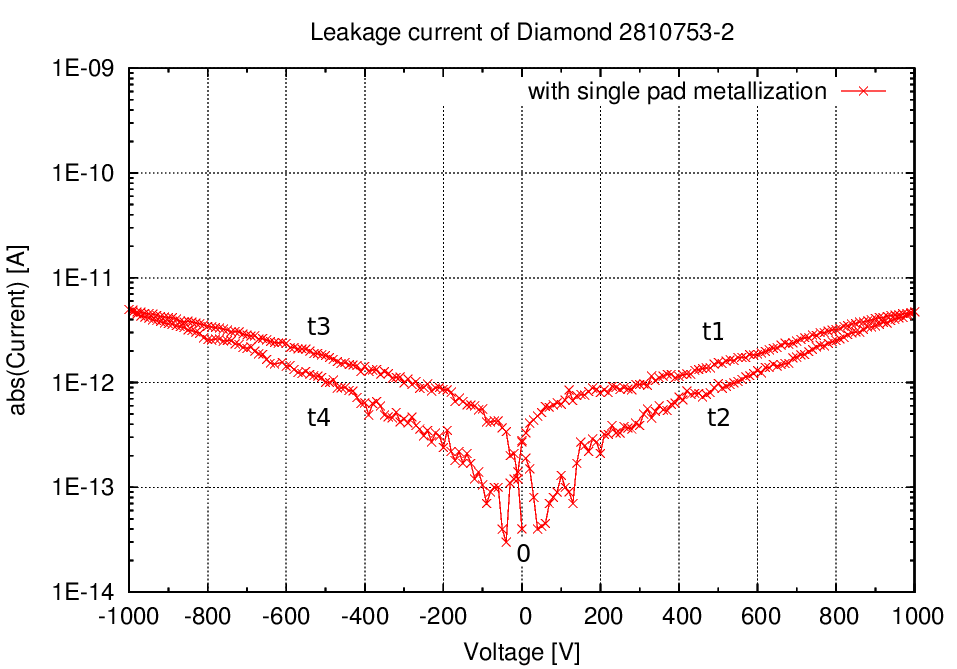}
 \caption{Leakage current versus voltage for a one pixel diamond plate. t1, t2, t3 and t4 in the Figure identify the order in time of the measurements of the four curves.}   
  \label{fig:IV-curves}
 \end{center} 
\end{figure}

Each diamond plate is characterized in a specifically built IV setup.
The current is measured for both polarities while increasing the HV to 1 kV in steps of 10\,V and then the time stability of the current is checked during a period of the order of one hour.
The leakage current versus voltage (IV curve) for one diamond and for both polarities are shown in Figure~\ref{fig:IV-curves}; t1, t2, t3 and t4 in the Figure identify the order in time of the measurements of the four curves.



 More than 80 \% of the TOTEM scCVD diamond metalized plates reached 1 kV with a leakage current of less than 1 nA; the ones with larger leakage current were discarded.

\begin{figure}[hbt]
  \begin{center}
\includegraphics[width=0.6\linewidth,trim={2cm 0 0 0.6cm},clip]{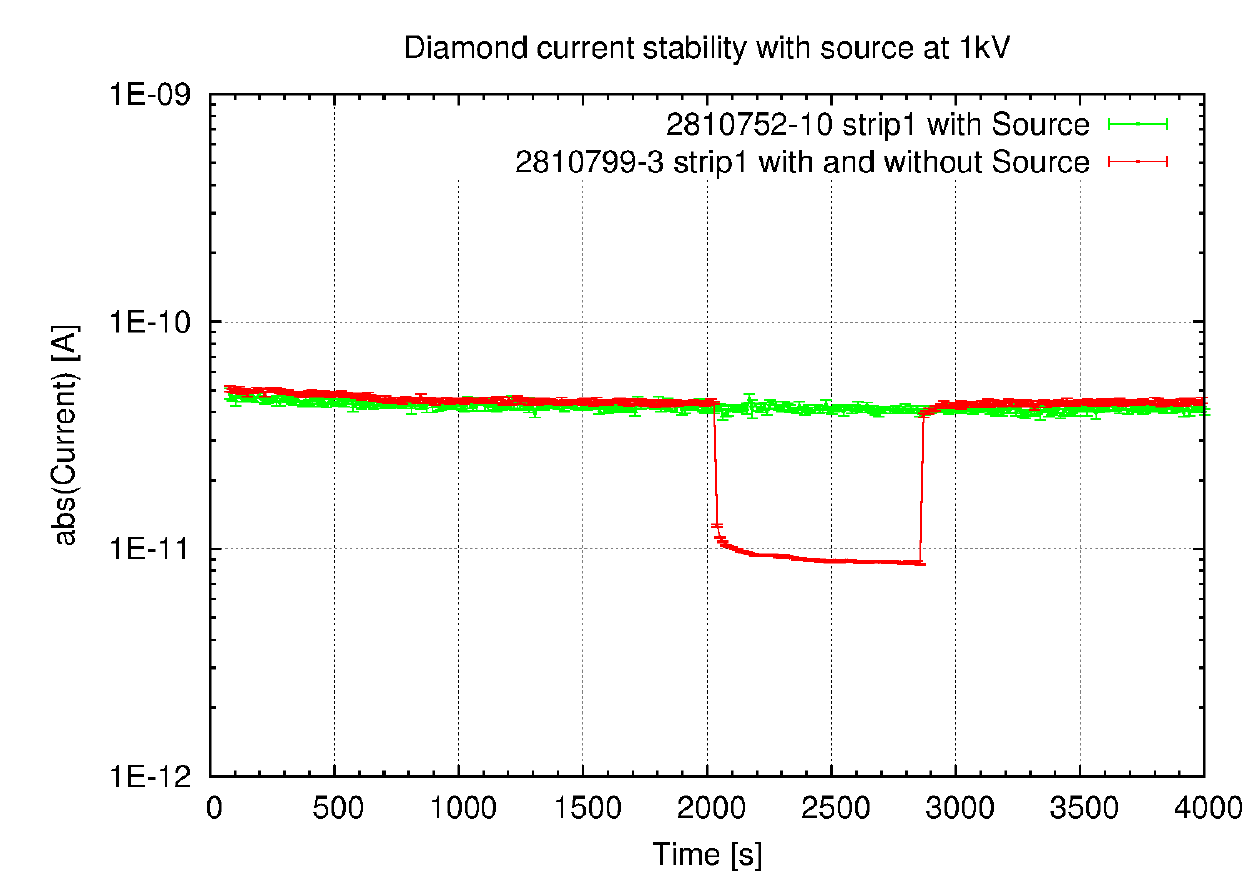}
 \caption{Stability of the signal currents over time for two different  metalized crystals biased at 1kV when exposed to a $^{90}$Sr radioactive $\beta$-source, which indicates the quality of both the metalization and the crystal itself.  }
  \label{fig:source-tests}
 \end{center} 
\end{figure}

The signal stability in time, an indicator of the metalization quality,  was measured using a $^{90}$Sr $\beta$ radioactive source of approximately 36\,MBq activity.
Tests of detectors with different metalization techniques and providers in identical conditions show similar performance in terms of charge collection and efficiency both with radioactive sources and with minimum ionizing particles see also section~\ref{sec:tests}.
Figure~\ref{fig:source-tests} shows the time stability of the signal current of two diamond detectors under irradiation from a radioactive $\beta$-source; the rate during these tests was approximately 10\,kHz per pixel. 
The red curve shows the current measurement with the radioactive  source removed at $\sim 2000$\,s and restored at $\sim 2800$\,s.

\subsection{Electronics}

Diamond sensors have the advantage of higher radiation hardness, lower noise and a faster charge collection than silicon sensors, however for equal thickness the amount of charge released by a Minimum Ionizing Particle (MIP) in the diamond is lower than for silicon or germanium: only $~$2.9 fC are released by a MIP in a 500 $\mu m$ diamond plate and a low noise amplifier is needed to keep the S/N ratio large enough.

The front end electronics design must be a compromise between speed and low noise.

Considering that the diamond resistivity is very high, the main source of noise will come from the first stage of the amplifier. 
For optimal energy transfer from the source to the input of the amplifier the impedances must match. 
That implies a high input impedance for the first stage of the amplifier, which is not the optimal choice from the Electromagnetic Compatibility (EMC) point of view. 

\subsubsection{The  Amplifier}

\begin{figure}[ht]
 \begin{center}
\includegraphics[width=0.7\linewidth]{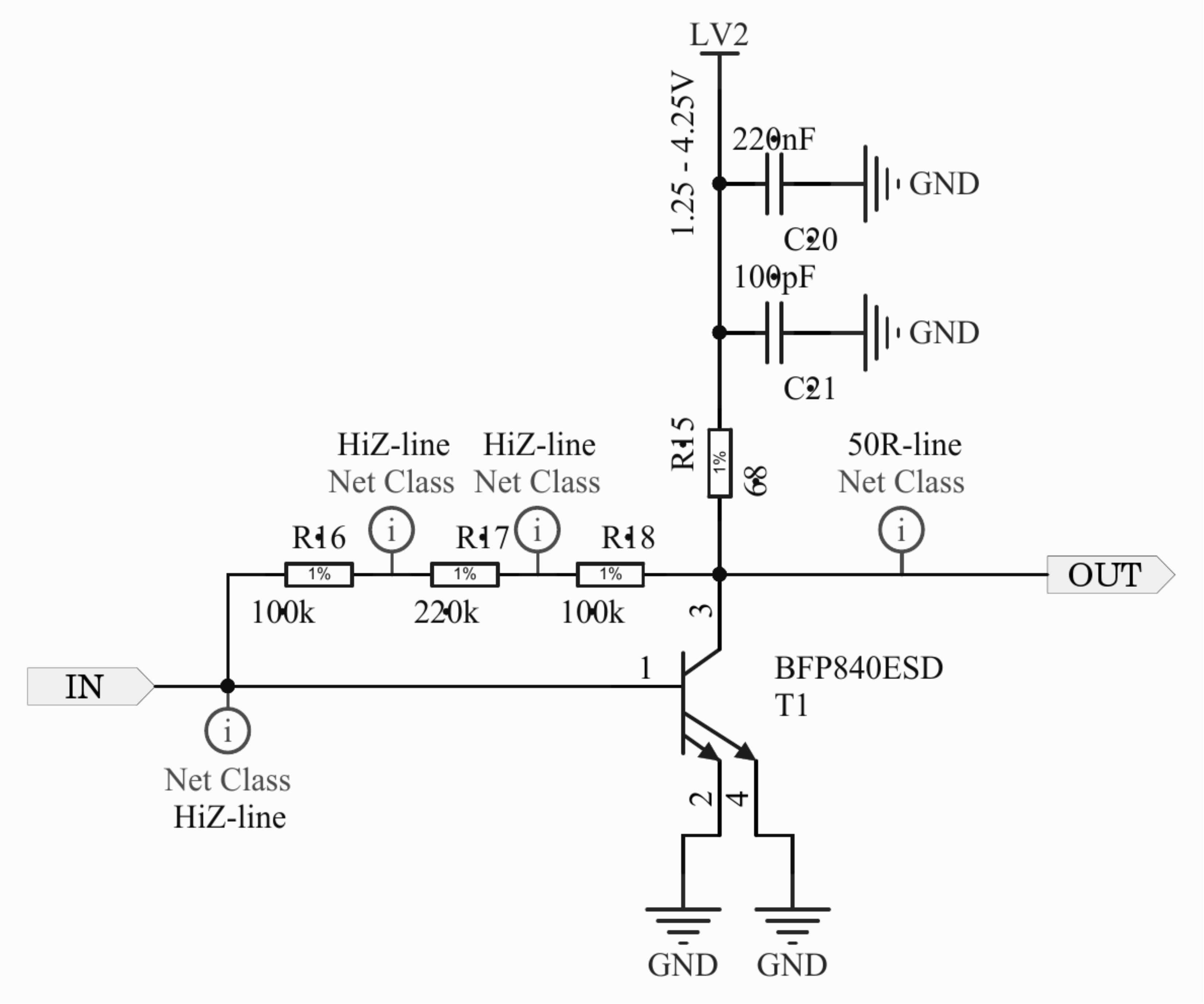}
\caption{The preamplifier stage of the TOTEM timing detector signal amplification chain.}
\label{schema}
\end{center}
\end{figure}

\begin{figure}[!]
   \begin{center}
   \includegraphics [width=.7\linewidth] {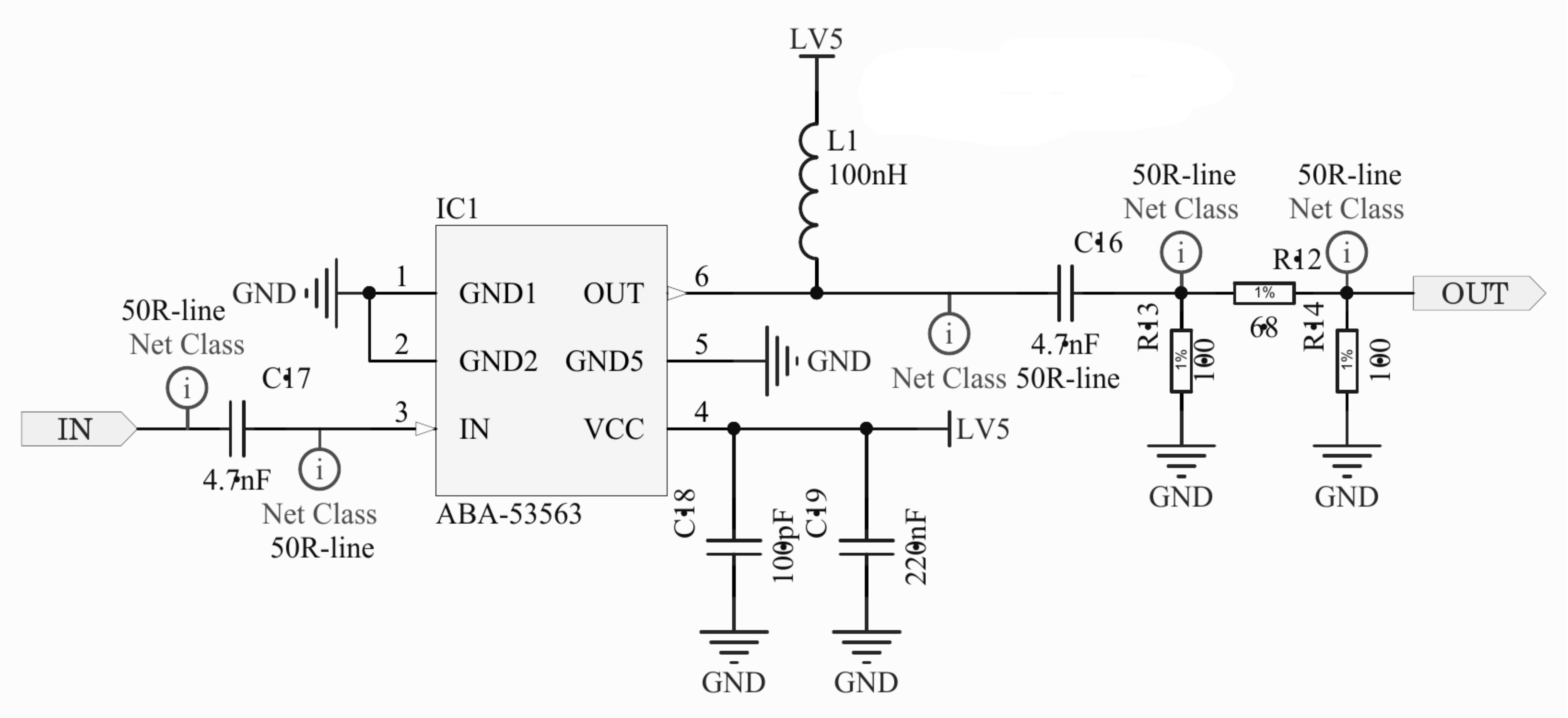}
   \end{center}
   \caption{The ABA stage of the TOTEM timing detector signal amplification chain.}
   \label{fig:ed_aba}
\end{figure}

\begin{figure}[!hbt]
   \begin{center}
   \includegraphics [width=.7\linewidth] {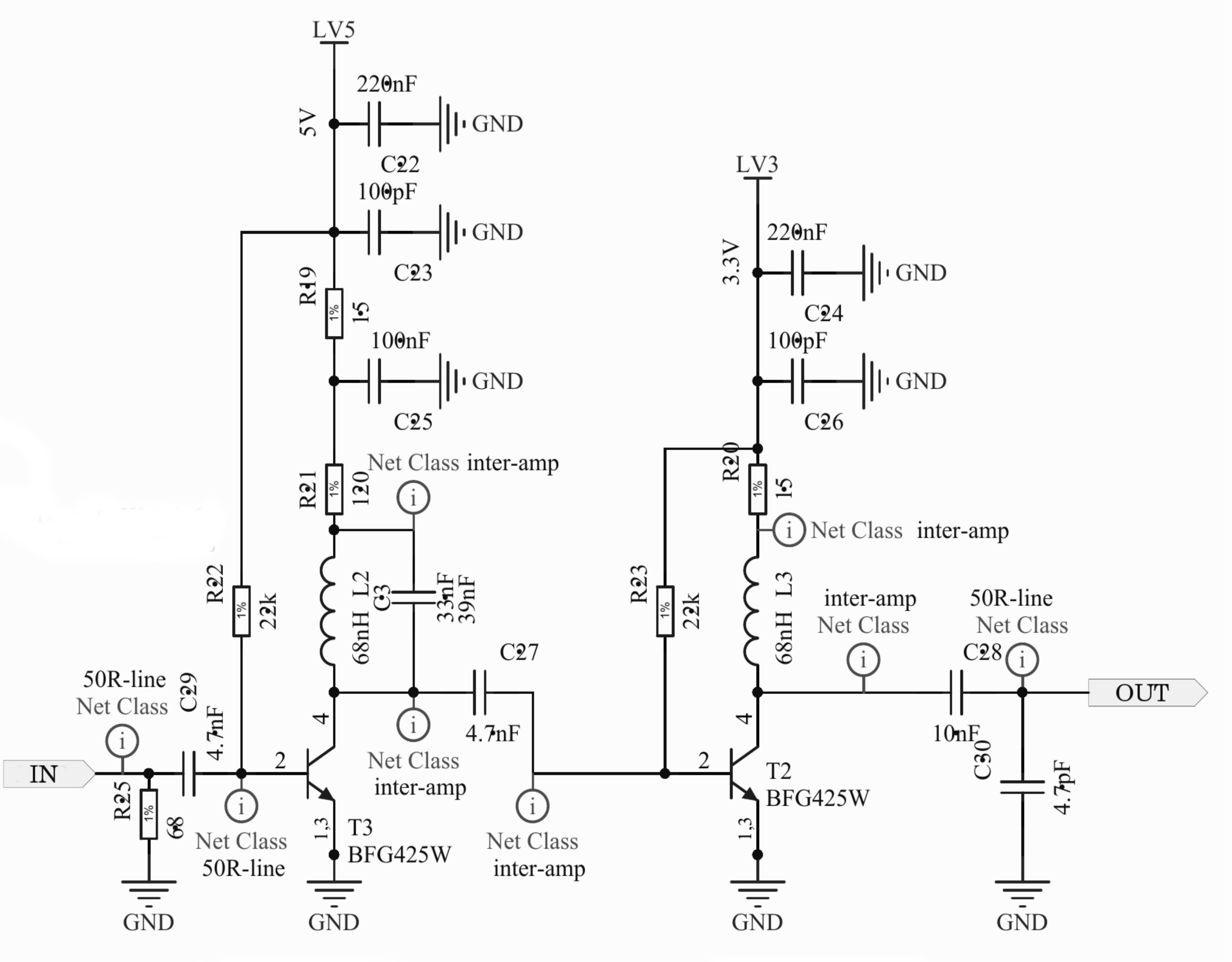}
   \end{center}
   \caption{The booster stage of the TOTEM timing detector signal amplification chain that both amplifies and shapes the signal. }
   \label{figure:booster_totem}
\end{figure}

The amplification chain designed for TOTEM consists of a transconductance preamplifier followed by a single stage voltage amplifier (ABA) and by a booster which also shapes the analogue voltage output signal.
The three parts of the amplification chain are shown in the Figures~\ref{schema}, \ref{fig:ed_aba} and \ref{figure:booster_totem}. 
This design has been adapted from the HADES Collaboration~\cite{Berdermann:2009eqa,5960819} amplifier. 

A good description for the diamond signal for the typical bias value used here is a triangular current pulse with a maximum amplitude of $\sim 1\, \mu A$, a rising edge of $\sim$\,2 ps and a falling edge of $\sim\,$5 ns. 
The preamplifier, a SiGe BJT with low-capacitance feedback in a common emitter configuration, see Figure~\ref{schema}, should be as close as possible to the sensor, at most 1 cm away.
The total input capacitance is given by the diamond capacitance (0.2 to 2 pF depending on the pixel size), by the connection of the diamond to the transistor base (approximately 0.2 pF when using 25 $\mu$m diameter wires) and by the parasitic capacitance of the transistor (about 0.4 pF). 
The input impedance for this configuration is a few k$\Omega$ while the capacitance seen by the signal source is low, which guarantees a high signal to noise ratio (S/N). 
The choice of the feedback and collector resistances of the preamplifier has been optimized by TOTEM in order to keep the time precision around 100 ps also for the electrodes with the largest capacitance ($\sim$2pF). 
The preamplifier has a maximum current gain of 45 dB at low frequency and of 93 dB at 200\,MHz.

The signal from the first stage is fed to the second stage (Figure~\ref{fig:ed_aba}) built around a broadband amplifier, the Avago ABA-53563 chip .

The optimization of the bias network of the ABA-53563 amplifier was performed in order to obtain an undistorted phase and gain response.
This circuit has an almost flat power gain over the frequency range of interest and a noise degradation factor of 3.5 dB.
The amplifier is suitably matched on input and output to 50\,$\Omega$ and is unconditionally stable. 
The main advantage of the ABA amplifier chip is a constant group delay which means a low shape distortion of the input signal.

An attenuator is inserted between this amplification stage and the booster stage to prevent saturation for the largest signals and to ensure a better matching of the two stages; the attenuator between the amplifier and the booster was set to 10\,dB  after tests for optimum performance.
The shaping stage is composed of two wide-band BJT transistors NPX BFG425W.  
The first booster stage defines the shaping time of the signal, via the RLC circuit between the collector and the feedback of the transistor. 
The shaper has a power gain of 53 dB at 200 MHz and an effective noise bandwidth of about 425 MHz.
The rise time from the last stage is about 1.7 ns and the max voltage (before saturation) is 2.7\,V, values that are adequate for transmission without degradation by approximately 2\,m of coaxial cable. 
For the specified input and supply voltage the RMS noise on the output  is $\sim\,$20 mV.

The schematic of the booster stage of the amplifier of the TOTEM timing detector is reproduced in Figure~\ref{figure:booster_totem}.


Table~\ref{table:electronics} summarizes the main parameters of the electronics.

\begin{table}[!]
\centering
\caption{Characteristics of the TOTEM amplifier.}
\begin{tabular}{lcc}

\hline
Number of stages  & \multicolumn{2}{l}{ 4 } \\
\hline
& \multicolumn{2}{l}{transconductance amplifier} \\
 &  \multicolumn{2}{l}{flat freq response amplifier (ABA)} \\
 & \multicolumn{2}{l}{10 dB attenuator} \\
 &\multicolumn{2}{l}{ booster-shaper} \\
 \hline
 & at DC & at 200 MHz \\
\hline
Input impedance & 10 kOhm  &1.5 kOhm \\
Current gain &  & 93 dB \\
Current gain 1$^{st}$ stage & 45 dB & 30 dB \\
Power gain 2$^{nd}$ stage &  \multicolumn{2}{c}{21 dB} \\
Attenuation 3$^{rd}$ stage &  \multicolumn{2}{c}{10 dB} \\
Power gain 4$^{th}$ stage &  & 52 dB \\
Power dissipated & 0.3 W/ch \\

\hline
\end{tabular}

\label{table:electronics}
\end{table}

\subsubsection{The Hybrid}

\begin{figure}[ht]
 \begin{center}
\includegraphics[width=.6\linewidth]{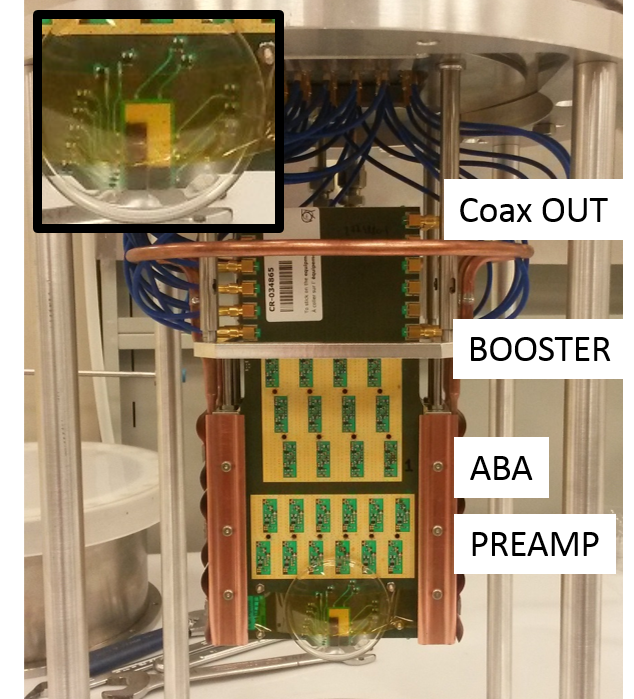}
\caption{Picture of one detector package in which the first of the four hybrid boards is visible: in the inset a close-up of the four diamond crystals. Starting from the bottom of the board, one can identify the three amplifications stages: the preamplifier stage (“PREAMP”), the ABA stage (“ABA”) and the booster stage (“BOOSTER”). On the top of the picture one can see the coaxial output cables (“Coax OUT”).}
\label{picture-hybrid}
\end{center}
\end{figure}

The hybrid board to accommodate both the electronics and the detectors is a four-layer board of Rogers RO4350B and RO4450F \footnote{ROGERS Corporation, 1 Technology Drive, Rogers, CT 06263, CT, USA.} low-loss materials ideal for RF applications and also a very good insulator for the HV bias of at least 700\,V required by the diamonds.
The detectors are positioned and aligned with a precision better than 0.1\,mm on this board and glued with silver-filled, electrically conductive epoxy adhesive\footnote{Ablestik Ablebond 84-LMI.}.
The picture of a timing detector plane designed for twelve channels hybrid and four diamond plates per hybrid is shown in Figure~\ref{picture-hybrid}, where the three hybrid amplification stages are indicated; the High Voltage (HV) pad onto which the diamonds are glued is easily identified. 

\subsubsection{Mechanical Challenges}

The detector packages (four hybrids) are in a vacuum of about $40$\,mbar that is required to minimize the deformation of the thin window of the RP that separates the detectors from the LHC circulating beam. 
The vacuum acts also as a heat insulator, and a heat exchanger is necessary to cool the detector and the RP itself.

The analog signals are sent from the hybrid inside the RP vacuum to the digitizer placed outside of the RP  vacuum vessel at a distance of 2\,m.
Special care has been taken to to minimize the reflections of the signal with a special flange equipped with 50\,$\Omega$ SMP coaxial vacuum feed-through\footnote{SMP standard connectors vacuum feed-through provided by \href{http://www.e6.com}{Solid Sealing Technology, Inc.} 44 Dalliba Avenue  Watervliet, New York 12189}.
Vacuum also imposes special care in routing the HV to the detectors as will be described later in section~\ref{sect:hvprotection}.

The four hybrids in one detector package are kept at a distance of 1\,cm from each other in a precision mechanical support that integrates also the cooling system. 
The support ensures the precise positioning of the edge of the sensor with respect to the  $200\, \mu$m thick `thin RP window': the presence of the thin window and the precise positioning of the detector near it are features of high importance for the physics measurement. 
Before the installation in the RP, the position of the edge of the diamond mounted  closest to the RP thin window, hence to the circulating beam, is measured with a laser with a precision of 10\,$\mu$m.
The alignment of the board on the support is then tuned to leave a safety margin of 150\,$\mu$m between the detector and the thin window.

\subsection{Digitization: the SAMPIC Chip}

The precise measurement  of the particle crossing time is obtained by digitizing the analog signals with a SAMpler for PICosecond time (SAMPIC) chip that gives a fast response similar to the one of a state-of-the-art TDC and the versatility of a waveform digitizer to perform accurate timing measurements.
The input bandwidth is 1.5 GHz,  the ADC precision is 11 bits and the unit may operate with a sampling frequency up to 10 GSa/s.

The  SAMPIC chip has been designed by a collaboration between CEA/IRFU/SEDI, Saclay and CNRS/LAL/SERDI, Orsay~\cite{Delagnes:2015oda,Breton:2016zoz} and named `Waveform and Time to Digital Converter' (WTDC) multichannel chip. 
Each channel associates a Delay Line Loop-based (DLL) TDC providing a raw time with an ultrafast analog memory allowing the extraction of precise timing and other parameters from the pulse.
Each channel also integrates a discriminator that can trigger itself independently or participate in a more complex trigger~\cite	{delagnes:in2p3-01082061}. 
After triggering, the analog voltages are digitized by an on-chip ADC and only the data corresponding to a region of interest are transferred serially to the DAQ. 
The association of the raw and fine timings permits to achieve a timing precision of a few ps RMS. 

Each channel is sampled through a 64 cell  Delay Locked Loop (DLL) based memory. 
When the acquisition is triggered, the sampling is stopped and the sampled values are digitized using a Wilkinson ADC.
The calibration procedure of the SAMPIC ADC is  automatically done by the control software sending to each input a ramp of known voltages, provided by accessory electronics on the mezzanine, and measuring the digitized value. 
A polynomial curve is used to compensate for non-linearities. 
The small variation of the internal length of the delay lines can be calibrated with an externally generated sine wave and by measuring the deformation of the digitized signal; a future version of the mezzanine will integrate a signal generator to automate this second calibration.


The time information is computed off-line from the digitized pulse using smart algorithms like constant fraction discrimination and fast cross-correlation.

A specifically designed TOTEM motherboard for synchronization of the acquired data with the TOTEM trigger and DAQ system is equipped with a mezzanine board with a SAMPIC chip mounted on it.

\subsection{Clock Distribution and DAQ Board}
 
To measure precisely on each side of the interaction point the arrival times of the protons and their difference, 
 it is necessary to send a common picosecond range precision time reference signal to the detectors. 

The TOTEM clock distribution design is adapted from the ``Universal Picosecond Timing System'', developed for FAIR at GSI ~\cite{PhysRevSTAB.12.042801}. 
The TOTEM clock distribution system, based on an optical network, uses dense wavelength division multiplex (DWDM) technique to transmit two reference clock signals from the counting room to a grid of receivers near the detectors. 
 The variation in the delay in the propagation of the signal is constantly measured in the central system by monitoring a signal reflected back from each station; this allows to control all the transmission systematic effects and keep the jitter below 5\,ps. 
A full prototype of the clock distribution system has been assembled and is under final tests.

The information from the TOF detectors are collected by a dedicated board, built around the radiation tolerant Microsemi Soc FPGA M2S150-FCG1152\footnote{ \href{http://www.microsemi.com}{Microsemi Corporation Corporate Headquarters}, One Enterprise Aliso Viejo, CA 92656 USA}. 
 The board receives the TOF signals digitized by the SAMPIC, builds the event data frame and sends them to the existing DAQ~\cite{Anelli:2008zza}.

The board also controls the SAMPIC and as an example, sets the thresholds of the channels. 
The expected rate of timing detector hits in the vertical RPs is expected to be $<$\,100\,kHz per plane, which can be handled by the SAMPIC digitizer.

\section{Tests and Detector Performance.}

In the process of designing the timing detectors, extensive tests have been performed with particle beams, initially to characterize the diamond sensors and to define the best design for the front end electronics and then to perform the check of the fully assembled prototype before its installation in the LHC. 

\subsection{Efficiency of the Segmented Diamond Detector.}

Efficiency measurements of a diamond sensor with four pixels (strips) have been performed with a prototype board on a mixed pion and electron beam at DESY where a tracking telescope was available.
To measure the efficiency the tracks have been reconstructed with the  DATURA tracking telescope\footnote{DESY Advanced Telescope Using Readout Acceleration built as part of the EUDET\textbackslash AIDA project.} composed of six silicon pixel detection planes.
\begin{figure}[!h]
   \begin{center}
   \includegraphics [scale=.55] {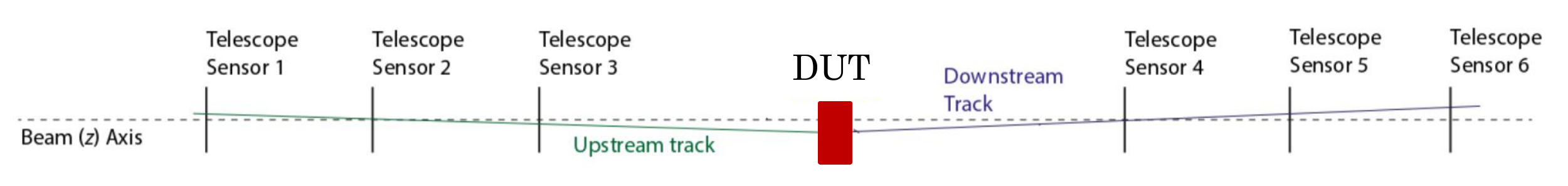}
   \end{center}
   \caption[Setup of the DATURA telescope]{Setup of the DATURA telescope. Tracks are independently reconstructed on each arm.}
   \label{figure:datura_concept}
\end{figure}
Figure~\ref{figure:datura_concept} shows the \emph{Device Under Test} (DUT) between two sets of three MIMOSA-26\footnote{https://twiki.cern.ch/twiki/bin/view/MimosaTelescope/WebHome} silicon pixel detectors.
These CMOS pixel sensors are 50 $\mu$m thick with an active area of $21.2\times10.6$ mm$^2$  subdivided into 1152 columns of 576 pixels each, i.e. a segmentation with a pitch of $\sim18.4$ $\mu$m, and a final precision	 of $\sim5$  $\mu$m for each plane.
The off-line algorithm determines the impact point of the particle in the diamond detector reconstructing the track independently in each arm of the telescope thus allowing to consider the effect of possible multiple scattering in the DUT. 
This feature is particularly useful when working with electrons. 

The measurements were performed with three hybrid boards, two equipped with single pixel diamonds and one with a segmented diamond.
The four pixel sensor  was installed between the two arms of the tracking telescope while the single pixel ones were placed downstream of the tracker.
The sensors have been described in detail in section~\ref{sec:diamond}.
The two single pixel sensors were provided by CIVIDEC\footnote{CIVIDEC Instrumentation GmbH, Schottengasse 3A/1/41,  A - 1010 Vienna, Austria.} and Element6.
The detectors tested had been metalized by CIVIDEC and by PRISM, the Princeton University laboratory. 
The diamond plates with four pixels foreseen for the vertical upgrade were metalized by the GSI, Darmstadt, Germany.\label{sec:tests} 

The efficiency is measured by comparing the position of a reconstructed track with the presence of a signal in one pixel. 
Given the high precision provided by the tracker, it was also possible to perform a two-dimensional scan of the diamond.

\begin{figure}[!ht]
   \begin{center}
   \includegraphics [scale=.20] {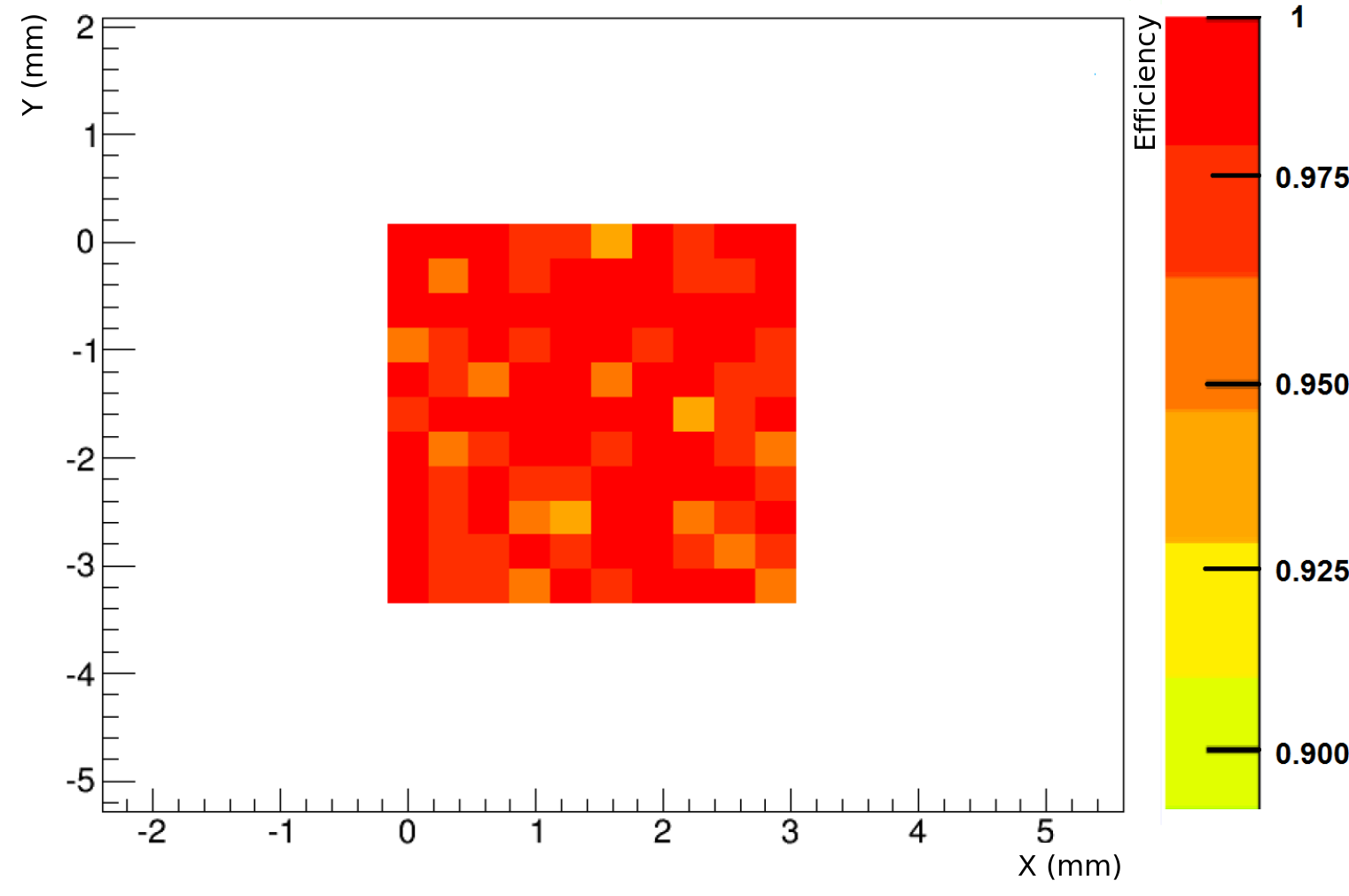}
   \end{center}
   \caption{The particle detection efficiency of the diamond detector. The statistical error on each point is $5-10$\%.}
   \label{figure:eff_global}
\end{figure}

A uniform efficiency above 98\% has been measured over a diamond detector, see Figure~\ref{figure:eff_global}. 

\begin{figure}[!t]
 \begin{center}
\includegraphics[width=0.8\linewidth]{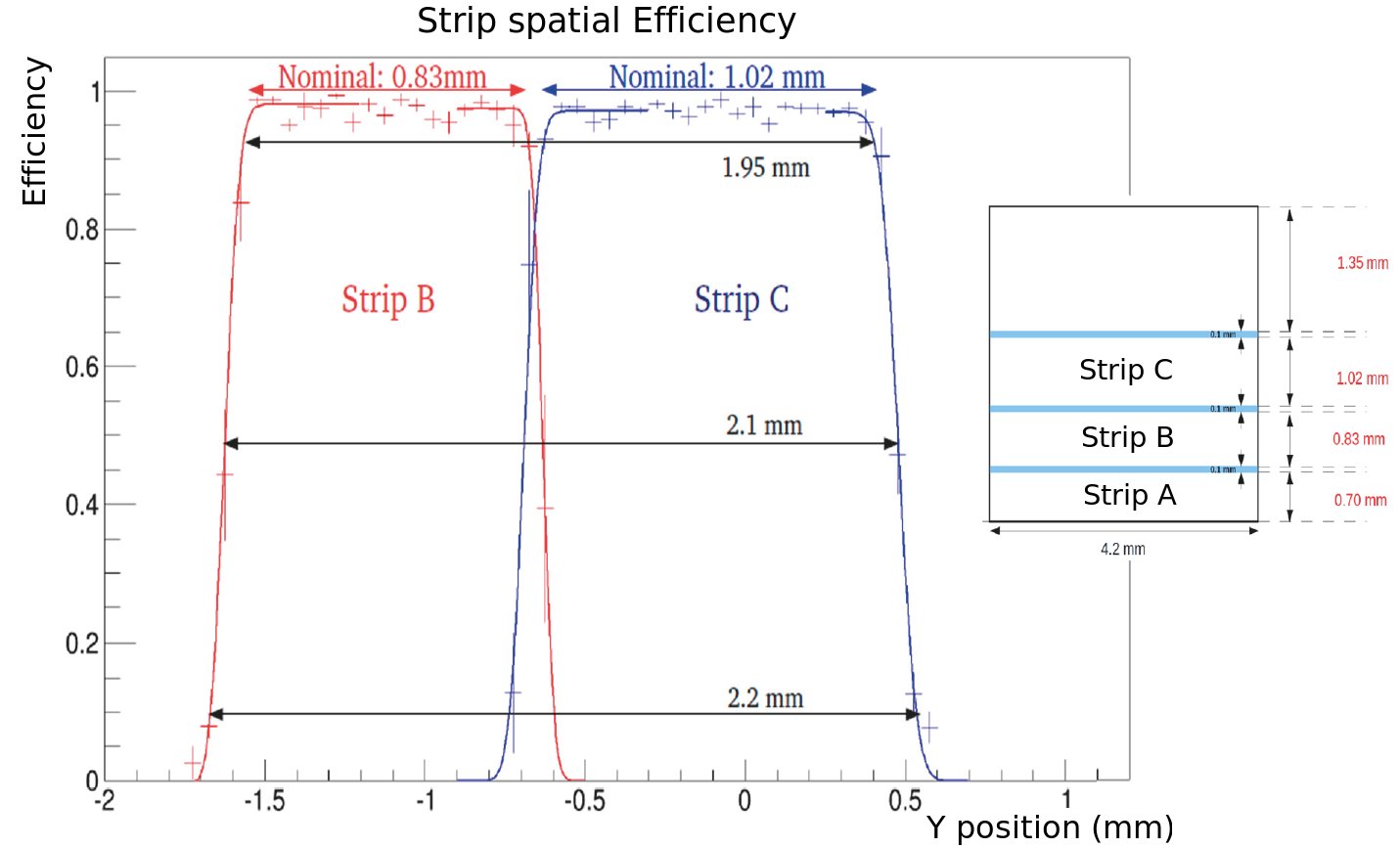}
\caption{Efficiency as a function of the position for two neighboring pixels separated by 0.1 mm. The position of the hit is defined by tracks reconstructed from silicon telescope data. A 1-2\% inefficiency can be attributed to the data acquisition system used for the test.}
\label{Effifig}
\end{center}
\end{figure}

The measurements of efficiency in the inter-pixel area are shown in Figure~\ref{Effifig}  for the projection of the efficiency on the y axis. 
Getting closer to the thin inter-pixel area (0.1 mm wide)\footnote{the geometrical region where the metalization is missing.},  the efficiency diminishes slowly while it starts  increasing for the adjacent pixel due to charge sharing.
It is worth noting that the efficiency of  one pixel remains always higher than 80\% in the inter-pixel area.
 
In the inter-pixel area we have therefore two pixels that are able to detect the particle,  and the overall combined efficiency becomes $\sim$96\%. 
Still, for the particles traversing the lower efficiency area, one expects a worse time precision, since the amount of charge collected from the pixel electrode is lower. 
With a  tracker detector associated to the timing detector, corrections based on the collected charge from both pixels can still be performed to reach the nominal precision.

\begin{figure}[!h]
   \begin{center}
   \includegraphics [width=.80\textwidth] {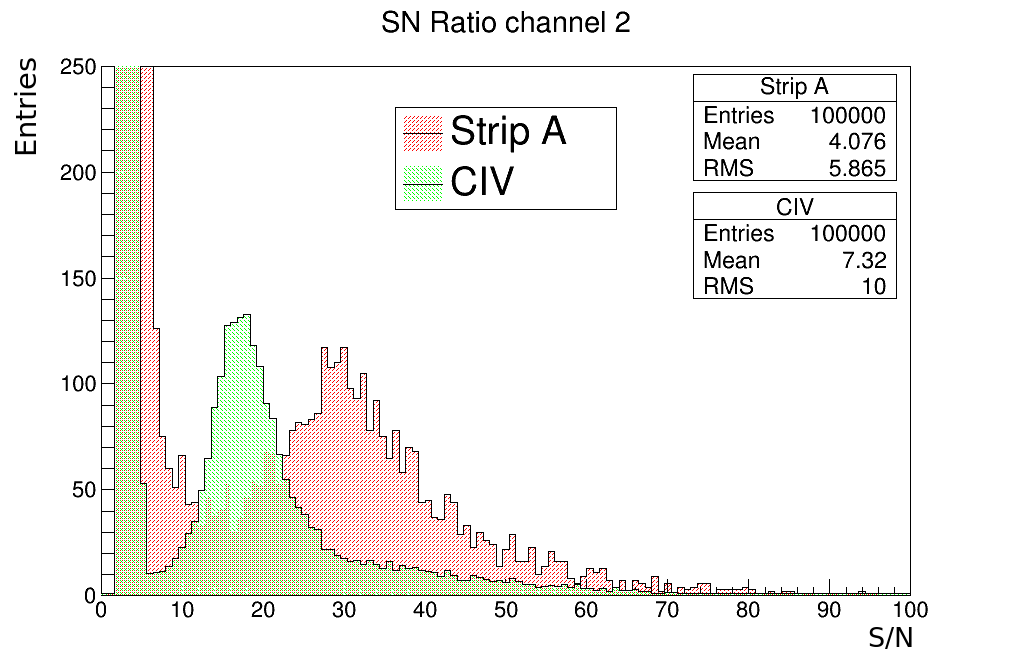} 
   \end{center}
   \caption{Comparison of the S/N of ``Strip A'' (a pixel with 0.29 pF) and CIVIDEC (2 pF).  The number of entries of each histogram are rescaled for better visualization of the signal region.}
   \label{figure:smallVsBig_pixel_par}
\end{figure}
Figure~\ref{figure:smallVsBig_pixel_par} shows, superimposed and appropriately rescaled for comparison, the S/N for the pixel with the smallest and largest capacitance.
The Signal to Noise ratio (S/N) obtained in these tests ranges between 20 and 35 with a small dependence on the pixel capacitance; the rise time of the pulse is about 1.7 ns.

\subsection{Time Precision}

Time precision measurements with two diamond boards equipped with the final electronics were performed first in a test beam of 5.6 GeV electrons at DESY, and repeated later in a 180 GeV $\pi^{+}$ beam at the CERN SPS.
The two tests gave consistent results.

\begin{figure}[!h]
   \begin{center}
   \includegraphics [scale=.20] {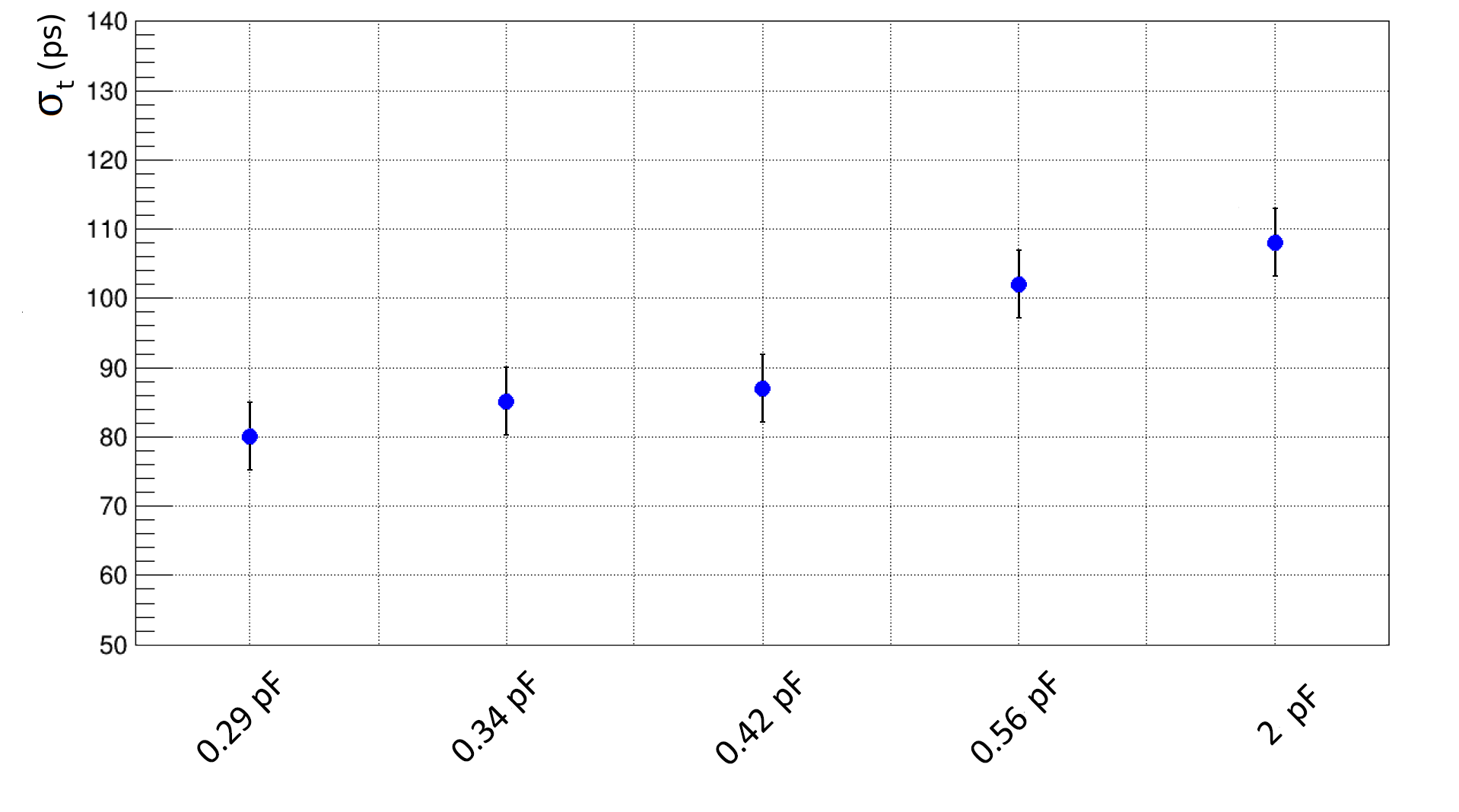}
   \end{center}
   \caption{Time precision as a function of the pixel capacitance and hence the detector area (after electronics optimization).}
   \label{figure:cap_dependence}
\end{figure}

 During these measurements performed at atmospheric pressure, the diamonds were biased at 700V.
The data acquisition was performed with an Agilent DSO9254A oscilloscope (8 bits, 20 GSa/s), and the waveforms were then analyzed off-line using different algorithms.
The time precision of a single detector is evaluated from the difference of the arrival time of a Minimum Ionizing Particle (MIP) traversing two diamond planes.
No external time reference was used, only signals from two hybrids, with all the channels equipped with the final electronics.
The time precision of the different pixels covers an interval between  $80 < \sigma_t < 108 $ ps and correlates with the pixel capacitance that  ranges between $0.29 < C < 2$ pF.
Figure~\ref{figure:cap_dependence} shows the time precision as a function of the pixel capacitance; these results are summarized in Table~\ref{table:cap_dep}.
 Pixels from different diamond sensors with the same capacitance show, within 5\,ps, a similar time precision .
 
\begin{table}[!h]
\centering
\caption{Summary of the results obtained for different sizes of the pixel sensitive area. Uncertainties of the order of 5\% were estimated by looking at the response of different electronic channels with pixels of the same capacitance.}
\begin{tabular}{c r c l p{1.2cm} c c c c c}
\hline
Sensor & \multicolumn{3}{c} {Pixel area}  & C& $\sigma_t$ & Ampl.& Rise Time & S/N\\
 & \multicolumn{3}{c} {[mm$^2$]}  & [pF] & & [V] &  [ns] &\\
\hline
\hline

Strip A&0.7 &$\times$&4.2 &0.29 & 80 &   0.74 & 1.48  & 30\\
Strip B&0.83&$\times$&4.2 & 0.34 & 85 &  0.70 & 1.51  & 31\\
Strip C&1.02&$\times$&4.2 & 0.42 &87 &  0.70 & 1.48  & 30\\
Strip D&1.35&$\times$&4.2 & 0.56 &102 & 0.62 & 1.49  & 28\\
One pixel  &4.2 &$\times$&4.2  & 2 & 108 &  0.39 & 1.56  & 18\\

\hline
\end{tabular}

\label{table:cap_dep}
\end{table}

\begin{figure}[!h]
   \begin{center}
   \includegraphics [scale=.25] {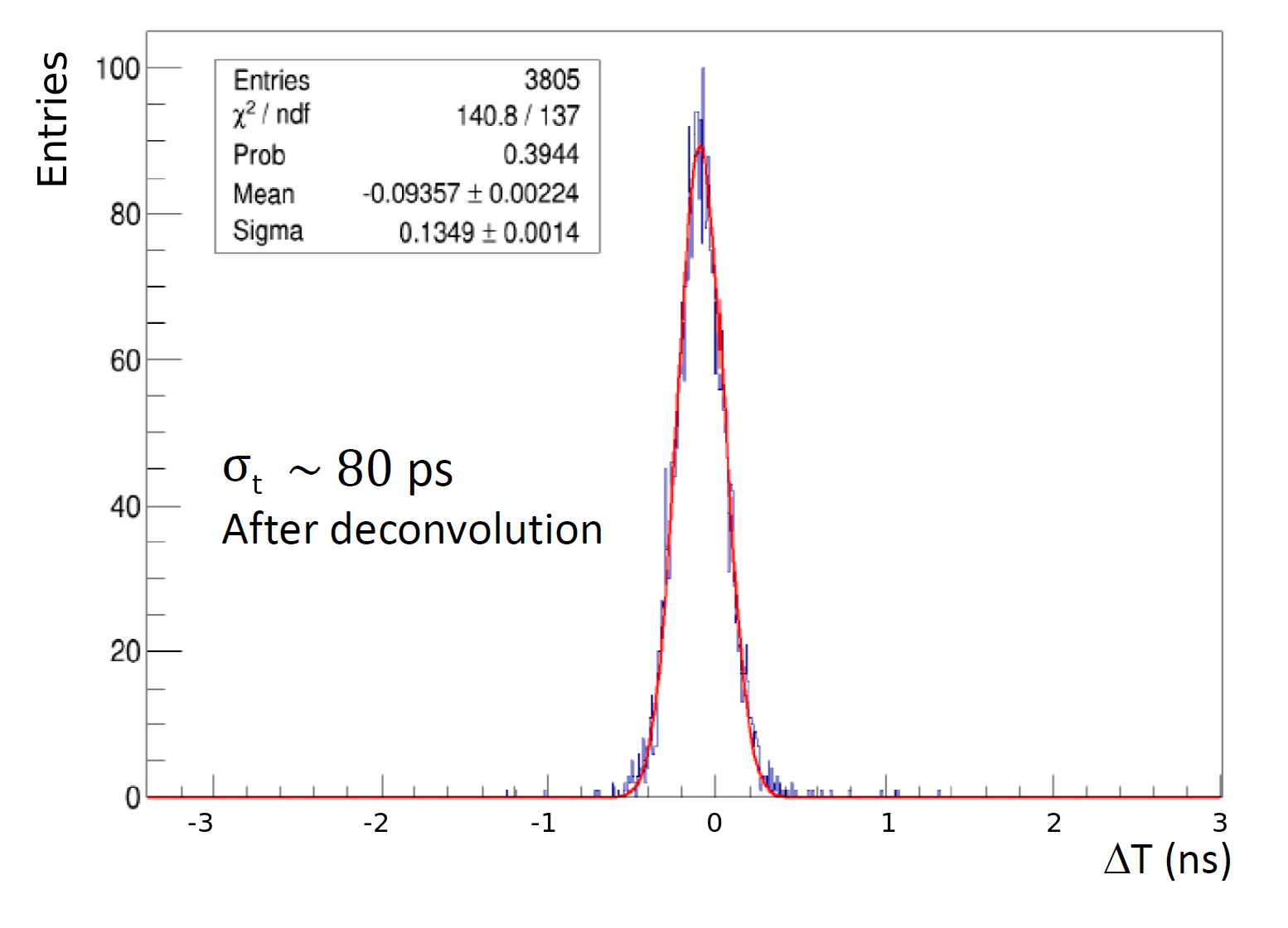}
   \end{center}
   \caption[Distribution of the time difference between a 2.0 pF and a 0.29 pF sensor]{Measured time difference between Strip A (0.29 pF) of a four pixel sensor and a one pixel (2.0 pF) sensor.}
   \label{figure:smallVsBig_pixel}
\end{figure}

Figure~\ref{figure:smallVsBig_pixel} shows the time difference distribution measured with the small pixel and the large pixel sensor.

The charge collection time, hence the peak current, depends on the drift velocity of the charge carriers. 
We observe that, due to non-saturation of the carriers' drift velocity in diamonds~\cite{pernegger.2005,Pomorski-thesis}, the time precision improves with increasing HV: tests have been repeated for different biasing voltages up to 700\,V or fields up to 1.4 V$\,/\mu$m. 
Above this value, the leakage current of some diamond sensors increased and the overall time precision did not improve further. 

The measurements reported in Figure~\ref{figure:cap_dependence} demonstrate that the $\sim100$ ps design precision needed for the project had been reached.
Measurements performed in similar conditions on a particle beam at CERN show that the time precision obtained using the SAMPIC digitizer is compatible within less than 5 ps with the one obtained with the high bandwidth oscilloscope. 

\section{Detector Package Integration and Installation}

The detector package is assembled from four hybrids mounted and aligned on the same mechanical structure.
The system must be able to work properly in the very specific environment of the TOTEM RP in the LHC. 
We describe now some technical system issues that imposed additional requirement on the detector.

\subsection{The HV distribution}\label{sect:hvprotection}

To measure protons very close to the LHC beam axis, detectors must be installed in RPs, which are under moderate vacuum.
Therefore special  attention was dedicated to the design of the HV distribution for the diamond crystals.

Systematic tests of final prototype hybrids under various vacuum conditions and for above-nominal operating HV values were performed to test the quality of the system under HV and vacuum.
When the onset of a discharge was seen from the current drawn by the HV power supply, a visual check allowed to identify the location where most likely the discharge had taken place.
To reduce the probability of HV discharges under the moderate vacuum operation conditions, a specific layer of the hybrid board (the $5^{th}$ layer) was solely dedicated to the routing of the HV distribution and the blocking capacitors have been positioned in regions free of other components.
The hybrid board, before adding the diamonds was tested to reach  at least 1200 V without discharges.
Once the diamonds were properly positioned, at 400 V one started to observe lateral discharges between the two faces of the diamond sensors.
This problem was solved as detailed in section~\ref{sec:coating}.

The pixels are connected to the input of the amplifier, approximately 1 cm away, with an  aluminum bonding wire of 150 $\mu$m diameter, which is thicker than the usual 25 $\mu$m diameter.
 Tests have also been performed to ensure that the small increase of capacitance and inductance due to a larger than usual wire diameter does not degrade the time precision of the detector.
 
 \subsection{The Coating}\label{sec:coating}
 On the hybrid board, the edges of each diamond sensor and all the components carrying HV near them have been specifically protected by adding a protective high volume resistivity Silicone coating compound\footnote{Dow Corning SE918L} .
 This operation had to be performed after the bonding and was studied in collaboration with the company Mipot\footnote{\href{http://mipot.com}{Mipot Hi-Tech  company S.p.A.}, Via Corona, 5, 34071 Cormons GO, Italy} that has then treated all our boards.
 
Final tests have been performed on all the eight boards treated with this feature and have shown that one could safely bring the HV to 700\,V at the required operating pressure of about 40\,mbar.

 \subsection{Vacuum Operation and Cooling}
The necessity to encase the detector package in a closed metallic container (the vacuum vessel) is a source of unwanted noise and feedback between the hybrids.
Measurements with a fully equipped detector package have shown a very large level of interference between the boards that induced easily  unwanted oscillations.
To damp any possibility of undesired radiative interference, the vacuum vessel is lined with 2\,mm EMI suppressor shielding from KEMET\footnote{\href{http://www.kemet.com}{KEMET Electronics Corporation, 2835 Kemet Way, Simpsonville, SC 29681, (USA)}.}.

To operate the electronics in the vacuum vessel all the dissipated  power must be removed by an efficient and stable cooling system.
In fact, the amplifier was designed for operation at room temperature.
The system must evacuate the power dissipated by the electronics and the heating of the RP vacuum pipe due to the beam RF pick up which increases at high luminosity.
The RP are equipped with an evaporative cooling system as already used for the silicon TOTEM strip detectors.
It has been set to keep the temperature stable at 25$^{\circ}$\,C.

 \subsection{Commissioning and Operation in the LHC Tunnel}
 
\begin{figure}[!t]
 \begin{center}
\includegraphics[width=0.65\linewidth,angle=-90]{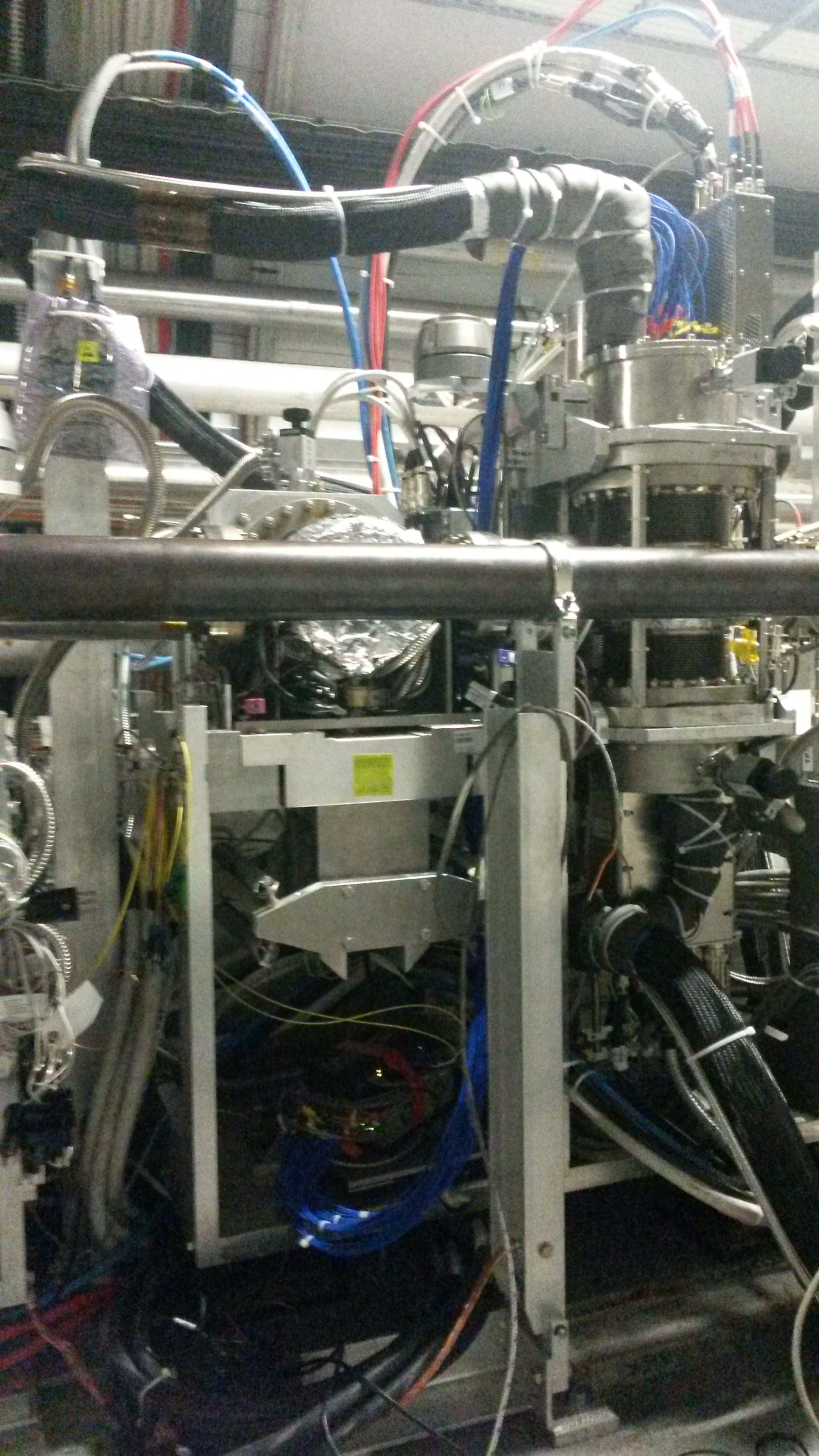}
\caption{The prototype TOTEM timing detector installed in the LHC tunnel in November 2015}
\label{fig:install--tunnel}
\end{center}
\end{figure}

The assembled system, installed in a Roman Pot container and cooled as if it were in the tunnel, was commissioned and checked on a particle test beam on the surface prior to the installation in the LHC tunnel in the November 2015 technical stop, see Figure~\ref{fig:install--tunnel}.

The timing detector was operated for 48 hours during an LHC $pp$ run at $\sqrt s$\,=\,5 TeV: the RPs were retracted in the `garage' position, and the observed signal rate was $\sim$10-100 Hz/pixel. 

\begin{figure}[!t]
 \begin{center}
\includegraphics[width=0.65\linewidth]{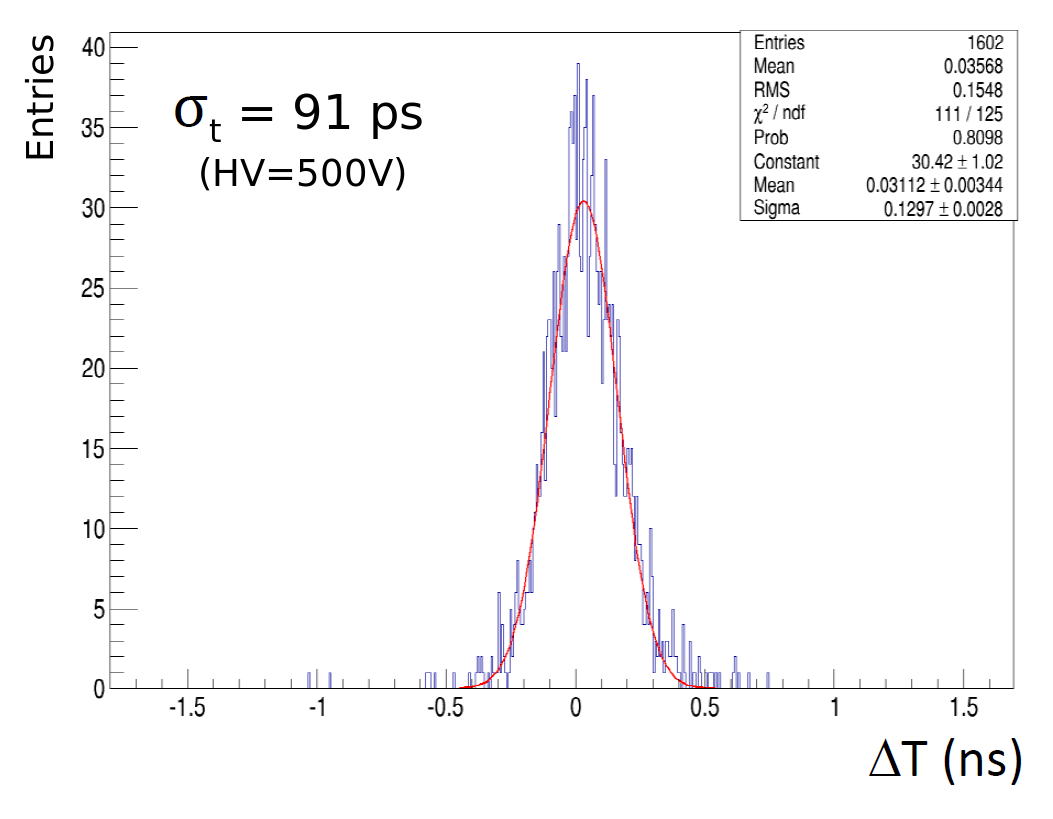}
\caption{Time precision for the prototype  as measured in the test LHC run in 2015.}
\label{fig:risol-tunnel}
\end{center}
\end{figure}

In the LHC, with the diamonds operated at an HV of 500\,V and with the digitization performed by a SAMPIC unit, the time precision obtained for the small capacitance pixels was $\sim$90 ps,  see Figure~\ref{fig:risol-tunnel}.
This is compatible with the 80 ps obtained in previous tests with a higher HV value. 

The noise of the electronic board at T\,=\,25$^{\circ}\,C$ was 15-28 mV RMS, compatible with the one previously measured. 

\section{Conclusions}

The  design of the TOTEM TOF diamond detectors for the upgrade of the vertical RPs has been described.
The sensors have been characterized in several test beams and have shown a time precision better than 100 ps per plane for MIPs, and for electrode capacitance less than~1\,pF. 
The detection efficiency is larger than 98\% in the metalized region. 
The digitization of the signal, made with the SAMPIC chip introduces a measured negligible deterioration of the time precision.
The first TOF prototype with four diamond planes read out by SAMPIC has been installed in the LHC at the beginning of November 2015. 
A time precision of about 90\,ps per plane has been found, compatible with the project specification. 
The clock distribution system, designed to guarantee a jitter control to better than 1 ps precision, has been assembled in the laboratory. 

Later,  in June 2016, two modules of four diamond planes have been installed in the LHC and are now operated at the highest LHC luminosity obtained so far ($1.4 \times 10^{34}\,{\rm cm}^{-2}\,{\rm s}^{-1}$) as part of the CT-PPS project~\cite{Albrow:1753795}.


\section*{Acknowledgments}

We thank S. Baud, D. Druzhkin and R. Stepanovich for the invaluable and efficient help during the development and installation of the detectors and F. Manolescu and I. Mcgill of the CERN bonding lab for their excellent bonding work on many hybrid boards and prototypes.
The efficiency  measurements reported in this work have been obtained at the test beam facility at DESY, a member of the Helmholtz Association (HGF). 
We are grateful to GSI and to PRISM (Princeton) for having provided help with the diamond metalizations. 
We thank the HADES colleagues for the useful discussions and for allowing us to test one of their detectors. 

This work was supported by the institutions listed on the front page and also by the Magnus Ehrnrooth foundation (Finland), the Waldemar von Frenckell foundation (Finland), the Academy of Finland, the Finnish Academy of Science and Letters (The Vilho, Yrj\"o and Kalle V\"ais\"al\"a Fund), the OTKA grant NK 101438 (Hungary) and also by the project LM2015058 from the Czech Ministry of Education Youth and Sports.


Individuals have received support from Nylands nation vid Helsingfors universitet (Finland), the Swedish Cultural Foundation in Finland and the M\v SMT \v CR (Czech Republic).

This project has received funding from the European Union's Horizon 2020~\footnote{https://ec.europa.eu/programmes/horizon2020/} Research and Innovation program under grant agreement No 654168.
Support for some of us to travel to CERN for the detector tests on the beam was provided by AIDA-2020-CERN-TB-2015-10 and AIDA-2020-CERN-TB-2016-11.

\bibliographystyle{unsrt}
\bibliography{Diam-2016}

\begin{thebibliography}{10}

\bibitem{Antchev:2011vs}
G.~Antchev, et~al.
\newblock {First measurement of the total proton-proton cross section at the
  LHC energy of $\sqrt{s} =7 $~TeV}.
\newblock {\em Europhys.Lett.}, 96:21002, 2011.

\bibitem{Antchev:2013iaa}
G.~Antchev et~al.
\newblock Luminosity-independent measurements of total, elastic and inelastic
  cross-sections at $\sqrt{s} = 7$~tev.
\newblock {\em Europhys.Lett.}, 101:21004, 2013.

\bibitem{totem4}
G.~Antchev et~al.
\newblock {Measurement of proton-proton elastic scattering and total
  cross-section at $\sqrt{s} = 7$~TeV}.
\newblock {\em Europhys.Lett.}, 101:21002, 2013.
\newblock CERN-PH-EP-2012-239.

\bibitem{PhysRevLett.111.012001}
G.~Antchev et~al.
\newblock Luminosity-independent measurement of the proton-proton total cross
  section at $\sqrt{s}=8\text{\,}\text{\,}\mathrm{TeV}$.
\newblock {\em Phys. Rev. Lett.}, 111:012001, Jul 2013.

\bibitem{totem5}
G.~Antchev et~al.
\newblock {Measurement of proton-proton inelastic scattering cross-section at
  $\sqrt{s} = 7$~TeV}.
\newblock {\em Europhys.Lett.}, 101:21003, 2013.

\bibitem{Ryskin:2007qx}
M.~G. Ryskin, et al.
\newblock {Soft diffraction at the LHC: A Partonic interpretation}.
\newblock {\em Eur. Phys. J.}, C54:199--217, 2008.

\bibitem{Gotsman:2008tr}
E.~Gotsman,  et~al.
\newblock {A QCD motivated model for soft interactions at high energies}.
\newblock {\em Eur. Phys. J.}, C57:689--709, 2008.

\bibitem{Ostapchenko:2011nk}
S.~Ostapchenko.
\newblock {On the model dependence of the relation between minimum-bias and
  inelastic proton-proton cross sections}.
\newblock {\em Phys. Lett.}, B703:588--592, 2011.

\bibitem{CERN-LHCC-2014-024}
{Addendum to the TOTEM TDR: Timing Measurements in the Vertical Roman Pots of
  the TOTEM Experiment LHCC document CERN-LHCC-2014-020 including
  questions/answers from/to the referees}.
\newblock Technical Report CERN-LHCC-2014-024. TOTEM-TDR-002-ADD-1, CERN,
  Geneva, Nov 2014.

\bibitem{Osterberg:2014mta}
Kenneth {\''{O}sterberg}.
\newblock {Potential of central exclusive production studies in high $\beta^*$
  runs at the LHC with CMS-TOTEM}.
\newblock {\em Int. J. Mod. Phys.}, A29(28):1446019, 2014.

\bibitem{CERN-LHCC-2014-020}
{Timing Measurements in the Vertical Roman Pots of the TOTEM Experiment}.
\newblock Technical Report CERN-LHCC-2014-020. TOTEM-TDR-002, CERN, Geneva, Sep
  2014.

\bibitem{TOTEM-TDR}
V~Berardi et~al.
\newblock Total cross-section, elastic scattering and diffraction dissociation
  at the large hadron collider at cern: Totem technical design report.
\newblock CERN-LHCC-2004-002 and addendum CERN-LHCC-2004-020, 2004.

\bibitem{Anelli:2008zza}
G.~Anelli et~al.
\newblock {The TOTEM Experiment at the CERN Large Hadron Collider}.
\newblock {\em JINST}, 3:S08007, 2008.

\bibitem{Ruggiero:2009zz}
G.~Ruggiero et~al.
\newblock {Characteristics of edgeless silicon detectors for the Roman Pots of
  the TOTEM experiment at the LHC}.
\newblock {\em Nucl.Instrum.Meth.}, A604:242--245, 2009.

\bibitem{Berretti:1747282}
M~Berretti.
\newblock {Performance studies of the Roman Pot timing detectors in the forward
  region of the IP5 at LHC}.
\newblock Aug 2014.

\bibitem{2007PSSAR.204.3004D}
W.~{de Boer},  et~al.
\newblock {Radiation hardness of diamond and silicon sensors compared}.
\newblock {\em Physica Status Solidi Applied Research}, 204:3004--3010,
  September 2007.

\bibitem{pernegger.2005}
H.~Pernegger,  et~al.
\newblock Charge-carrier properties in synthetic single-crystal diamond
  measured with the transient-current technique.
\newblock {\em Journal of Applied Physics}, 97(7):073704, 2005.

\bibitem{Pomorski-thesis}
M.Pomorski.
\newblock Electronic properties of single crystal cvd diamond and its suitability for particle detection in hadron physics experiments.
\newblock {\em PhD Thesis - Wolfgang von Goethe University Frankfurt - Aug. 2008. \\{\rm http://www-norhdia.gsi.de/\-publications/\-pomorski\_thesis\_final\-\_LQ.pdf}},
  2008.

\bibitem{Guthoff:2013hja}
Moritz Guthoff,  et~al.
\newblock {Radiation damage in the diamond based beam condition monitors of the
  CMS experiment at the Large Hadron Collider (LHC) at CERN}.
\newblock {\em Nucl. Instrum. Meth.}, A730:168--173, 2013.

\bibitem{Berdermann:2009eqa}
E.~Berdermann,  et~al.
\newblock {Diamond Start Detectors}.
\newblock In {\em {Proceedings, 2009 IEEE Nuclear Science Symposium and Medical Imaging Conference (NSS/MIC 2009)}}, pages 407--411, 2009.

\bibitem{5960819}
M.~Ciobanu,  et~al.
\newblock In-beam diamond start detectors.
\newblock {\em IEEE Transactions on Nuclear Science}, 58(4):2073--2083, Aug. 2011.

\bibitem{Delagnes:2015oda}
E.~Delagnes,  et~al.
\newblock {Reaching a few picosecond timing precision with the 16-channel
  digitizer and timestamper SAMPIC ASIC}.
\newblock {\em NIM A}, A787:245--249, 2015.

\bibitem{Breton:2016zoz}
D.~Breton,  et~al.
\newblock {Measurements of timing resolution of ultra-fast silicon detectors
  with the SAMPIC WTDC}.
\newblock {arXiv 1604.02385 - 2016}.

\bibitem{delagnes:in2p3-01082061}
E.~Delagnes et~al.
\newblock {The SAMPIC Waveform and Time to Digital Converter}.
\newblock {\em {Proceedings of the 2014 IEEE Nuclear Science Symposium and
  Medical Imaging Conference (2014 NSS/MIC), and 21st Symposium on
  Room-Temperature Semiconductor X-Ray and Gamma-Ray Detectors}}, November
  2014.
\newblock Sce Electronique.

\bibitem{PhysRevSTAB.12.042801}
M.~Bousonville  et~al.
\newblock Universal picosecond timing system for the facility for antiproton
  and ion research.
\newblock {\em Phys. Rev. ST Accel. Beams}, 12:042801, Apr 2009.

\bibitem{Albrow:1753795}
M~Albrow,  et~al. and the CMS-TOTEM Collaboration.
\newblock {CMS-TOTEM Precision Proton Spectrometer}.
\newblock Technical Report CERN-LHCC-2014-021. TOTEM-TDR-003. CMS-TDR-13, Sep
  2014.

\end{thebibliography}
\end{document}